\documentclass[aps,prd,twocolumn,preprintnumbers,nofootinbib,superscriptaddress]{revtex4-1}

\usepackage[colorlinks=true,citecolor=blue,linkcolor=blue]{hyperref}
\usepackage[normalem]{ulem}
\usepackage{amsmath,amssymb}
\usepackage{mathtools}
\usepackage{verbatim}
\usepackage{titlesec}               
\usepackage{epsfig}
\usepackage{graphicx}               
\usepackage{url}
\usepackage{color}
\usepackage{multirow}
\usepackage{floatrow}
\usepackage{placeins}
\usepackage[dvipsnames]{xcolor}
\usepackage{epstopdf}
\usepackage{fontawesome}
\usepackage{tikz}
\usepackage{tikz-feynman}
\usepackage{enumitem}
\usepackage[capitalize]{cleveref}
\usepackage{lipsum}
\usepackage{gensymb}
\usepackage{booktabs}               
\usepackage{xspace}                 
\usepackage{pifont}
\usepackage{marvosym }
\usetikzlibrary{shapes,arrows}
\usetikzlibrary{decorations.pathmorphing,decorations.markings}
\usetikzlibrary{snakes}
\usepackage{orcidlink}  
\usepackage{footnote}
\usepackage{xcolor}
\usepackage{fontawesome}
\allowdisplaybreaks

\usepackage{bbm}
\usepackage{slashed}

\makeatletter

\renewcommand{\p@subsection}{}
\makeatother

\titleformat*{\section}{\centering\bfseries\uppercase}
\titlelabel{\thetitle\quad}
\titleformat*{\paragraph}{\bfseries}
\titlespacing*{\paragraph}{0pt}{3.25ex plus 1ex minus .2ex}{1em}

\makeatletter
\def\l@subsubsection#1#2{}
\makeatother

\newsavebox{\twosubbox}

\begin{document}

\title{A Narrow Neutrino Window for the LZ Event}

\author{Vedran Brdar~\orcidlink{0000-0001-7027-5104}}
\email{vedran.brdar@okstate.edu}
\affiliation{Department of Physics, Oklahoma State University, Stillwater, OK 74078, USA}

\author{Dibya~S.~Chattopadhyay~\orcidlink{0000-0003-2323-3950}}
\email{dibya.chattopadhyay@okstate.edu}
\affiliation{Department of Physics, Oklahoma State University, Stillwater, OK 74078, USA}

\begin{abstract}
The LUX-ZEPLIN (LZ) experiment has reported an event consistent with a $248$ keV nuclear recoil. Explaining the absence of lower-energy recoil events typically calls for some form of upscattering that kinematically forbids such events. In this work, we present a framework in which the LZ observation has a neutrino origin, with atmospheric neutrinos providing the dominant flux in the required energy range. A scenario in which atmospheric neutrinos upscatter to a heavier neutral state would also produce a large number of neutral-current events in neutrino experiments. In particular, scattering on lighter nuclear targets results in much larger nuclear recoil energies compared to xenon, yet no such excess has been observed. We show that this constraint from neutrino experiments can be evaded if atmospheric neutrinos within a narrow energy window first produce a nearly monoenergetic state $N_1$, followed by the upscattering of $N_1$ to its heavier partner $N_2$ in LZ. In such a scenario, scattering on xenon becomes kinematically allowed for $N_2$ masses around $250$ MeV, while scattering on oxygen, carbon, and other targets used in large-scale neutrino experiments remains kinematically forbidden. We show that this two-step process, $\nu \to N_1 \to N_2$, can be realized through a parametric resonance induced by a dark matter background that efficiently produces $N_1$, followed by $N_1 \to N_2$ upscattering mediated by a vector boson in a model with gauged $U(1)_B$. The latter interaction can be sufficiently stronger than the weak interaction, which is necessary to lift the neutrino floor and yield $\mathcal{O}(1)$ event at LZ.
\end{abstract}

\maketitle

\section{Introduction}
\noindent
The LUX--ZEPLIN (LZ) Collaboration recently extended its nuclear-recoil search to energies of approximately $270$~keV and reported one event consistent with a recoil of $248\pm23\,(\mathrm{stat})\pm23\,(\mathrm{sys})$~keV in a $2.84$ tonne-year exposure~\cite{LZ:2026axp}. The event has a local significance of $3.4\sigma$, with a global significance of $2.6\sigma$. Its unusually high recoil energy has motivated interpretations chiefly based on inelastic scattering of dark matter (DM) near a kinematic threshold~\cite{Yin:2026jnn,Fan:2026kxx,Wu:2026nhi,Tucker-Smith:2001myb,Su:2026rwz,Freese:2026sga,Yamashita:2026ump,Pospelov:2026ewn,McCabe:2026crm,Rodd:2026tyn,Smirnov:2026aqk,Okada:2026upm,DiMauro:2026ldr,Dent:2026bji,Bose:2026ndd}.\footnote{See Ref.~\cite{CyncynatesLZMap} for a map of the steadily expanding list of interpretations and follow-up studies of the LZ high-recoil event.
}

In this work, we take a complementary approach and investigate whether the observed LZ event could have a neutrino origin. Applying the DM scattering approach mentioned above to neutrinos, the minimal scenario features the upscattering of an atmospheric neutrino into a massive neutral state $N$ through $\nu\,\text{Xe}\to N\,\text{Xe}$ \cite{Jeesun:2026vzo}. Neutrinos, however, are relativistic, and this qualitatively changes the situation relative to inelastic DM scattering. Namely, for a broad atmospheric neutrino spectrum, considering large-scale neutrino experiments with targets (such as oxygen, carbon, and hydrogen) lighter than xenon does not generally kinematically forbid the $\nu A\to N A$ process, where $A$ denotes the target nucleus. In fact, the nuclear recoil energy increases for lighter target nuclei. Since explaining the observed LZ event with neutrinos requires an interaction strength exceeding that of the weak interaction, the $\nu A\to N A$ process would therefore yield a large number of new-physics-induced neutral-current events in experiments such as JUNO~\cite{JUNO:2015zny,JUNO:2021vlw}, Super-Kamiokande~\cite{Super-Kamiokande:2002weg}, and IceCube~\cite{IceCube:2016zyt}. Given that such events have not been reported, the main challenge is to find a realization that strongly suppresses or completely eliminates interactions with lighter targets while preserving scattering off xenon.

\begin{figure}[t!]
    \centering
    \includegraphics[width=\linewidth]{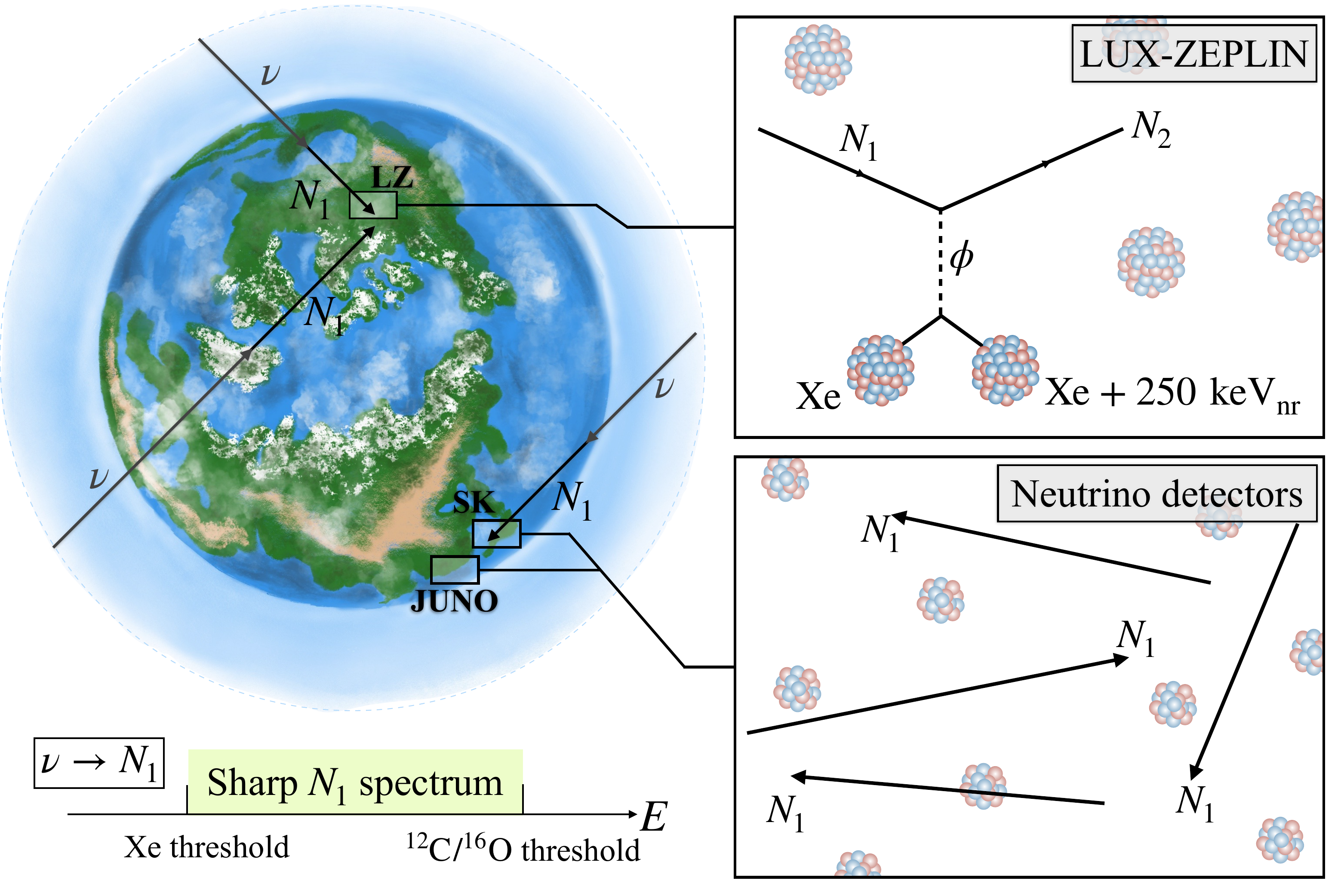}
    \caption{Artist's rendition of the proposed mechanism for neutrino-induced upscattering at LZ. Atmospheric neutrinos produce a narrow spectrum of $N_1$ that lies above the xenon threshold but below the carbon and oxygen thresholds. The upscattering process $N_1 {\rm Xe}\to N_2{\rm Xe}$ can then yield the observed recoil at LZ, while the corresponding channels in large neutrino detectors remain kinematically closed.}
    \label{fig:schematic}
\end{figure}

In this work, we overcome this hurdle and show that neutrino-induced upscattering can explain the LZ event. Namely, we show that the above argument has a kinematic loophole: scattering off low-mass nuclei in neutrino experiments can be kinematically forbidden if the scattering is instead initiated by a much narrower, nearly monoenergetic flux of a new neutral state.
In such a nearly monoenergetic scenario, the energy of this state can lie above the threshold energy for scattering off xenon but below the corresponding thresholds for scattering off lighter target nuclei. We note that, this threshold-based mechanism has recently also been applied to explain the LZ event in the context of upscattering in the boosted DM realization~\cite{Alhazmi:2026efz}.

In order to achieve such a narrow flux, we introduce an extended hidden sector containing a neutral state, $N_1$, and a DM background. In our setup, atmospheric neutrinos produce a narrowly peaked $N_1$ flux through a parametric resonance induced by the DM field. The resulting nearly monoenergetic $N_1$ states can then upscatter to a heavier neutral state $N_2$ in the LZ detector. The mass of $N_2$ can be such that the energy of $N_1$ lies above the threshold for scattering off xenon but below the corresponding thresholds for lighter nuclei.

The mechanism, which is also illustrated schematically in \cref{fig:schematic}, can be summarized as  
\begin{equation}
\nu\longrightarrow N_1 \,, \qquad N_1\, {\rm Xe}\longrightarrow N_2\, {\rm Xe} \,.
    \label{eq:chain}
\end{equation}

The paper is organized as follows. In \cref{sec:nu-N+kinematics}, we demonstrate that an atmospheric-neutrino upscattering explanation of the LZ event would also lead to a large number of neutral-current events at neutrino experiments that have not been observed. In \cref{sec:N1-N2}, we show that a minimal realization for explaining LZ and avoiding accompanying events in neutrino experiments features a nearly monoenergetic heavy neutral lepton, $N_1$, which upscatters to a heavier state $N_2$. In \cref{sec:narrow}, we demonstrate how such a narrow $N_1$ spectrum can be generated from atmospheric neutrinos, considering both matter-induced and parametric resonance. In \cref{sec:N1-N2-xsec}, we realize the $N_1\,\mathrm{Xe}\to N_2\,\mathrm{Xe}$ scattering process within a gauged $U(1)_B$ model and  identify regions of parameter space that are consistent with existing constraints and predict $\mathcal{O}(1)$ event at LZ. Finally, in \cref{sec:conclusion}, we conclude.

\section{Towards a Minimal Realization}
\label{sec:nu-N+kinematics}
\noindent
In order to forbid low-energy nuclear recoils at LZ that have not been observed in association with the 248 keV recoil event, the simplest neutrino-based explanation to assess is atmospheric neutrino upscattering to a heavier neutral state $N$~\cite{Brdar:2018qqj,Chao:2021bvq}. This~$\nu \,\text{Xe} \to N \,\text{Xe}$ process was proposed in Ref.~\cite{Jeesun:2026vzo}. The number of predicted events at LZ above recoil energy $T_0$ reads
\begin{align}
    N_{\rm LZ} = \mathcal E_\text{Xe,LZ} \int_{T_0} dT \int dE \, \Phi_\nu(E_\nu) \, \frac{d\sigma_{\text{Xe}}^{\nu N}(T,E_\nu)}{dT} \, \epsilon_{\rm LZ} (T)\, .
    \label{eq:direct-rate}
\end{align}
Here, $\mathcal E_\text{Xe,LZ}$ is the exposure, including the number of target nuclei and the data-taking time; $\Phi_\nu$ is the neutrino flux, which across the energies of interest is heavily dominated by atmospheric neutrinos, $T$ is the nuclear recoil energy, $d\sigma_{\text{Xe}}^{\nu N}/dT$ is the differential cross section, and $\epsilon_{\rm LZ}$ is the detection efficiency.

The process $\nu A\to N A$ is in general also realizable in neutrino experiments.
Broadly speaking, the comparison between the number of $\nu \, \text{Xe}\to N\,\text{Xe}$ events in LZ and corresponding events in a large neutrino detector is set by the target exposure, the atmospheric neutrino flux, the cross section, and detection efficiencies.

The exposure is, however, much larger for neutrino detectors. As an example, let us consider the JUNO experiment \cite{JUNO:2015zny,JUNO:2021vlw}. One year of data taking with its $\sim 20$ ktonne detector, with a $12\%$ hydrogen mass fraction, yields approximately
\begin{equation}
    \mathcal E_{p,{\rm JUNO}} \simeq 10^5 \, \mathcal E_{{\rm Xe},{\rm LZ}}\,,
\end{equation}
where $\mathcal E_{{\rm Xe},{\rm LZ}}$ corresponds to $2.84\,{\rm tonne\text{-}year}$ exposure \cite{LZ:2026axp,JUNO:2015zny,JUNO:2024jaw}. This implies that, for a single atmospheric neutrino scattering event detected at LZ, large neutrino detectors such as JUNO would detect orders of magnitude more neutral-current $\nu \,{\rm p} \to N \,{\rm p}$ events. However, before making such a claim, one needs to compare the detectable recoil-energy ranges at JUNO and LZ.

The threshold neutrino energy for producing a particle of mass $M_N$ through the process
$\nu \,\text{Xe} \to N \, \text{Xe}$ reads \cite{Brdar:2018qqj}
\begin{align}
E_\text{th}^\text{Xe}=M_N+M_N^2/(2m_\text{Xe})\,.
\label{eq:nuN1}
\end{align}
Taking $M_N=0.25$ GeV and $m_\text{Xe}=123$ GeV, this relation gives $E_\text{th}^\text{Xe} \simeq 0.25$ GeV. The corresponding recoil energy is
\begin{align}
T_0^{\text{Xe}}=M_N^2/(2m_\text{Xe}+2 M_N)\,,
\label{eq:nuN2}
\end{align}
giving $T_0^{\text{Xe}}\approx 250$~keV for the chosen parameters, implying that this benchmark point can in principle explain the observed recoil at LZ.

The JUNO collaboration, in the context of supernova searches, pointed out that proton recoil energies above $0.2$ MeV are expected to be observable~\cite{JUNO:2015zny}. More conservatively, in Ref.~\cite{Chauhan:2021fzu}, the authors discuss a visible scintillation energy above $15$ MeV, and we take that energy as a conservative benchmark. Using Birks' law~\cite{Birks:1951boa}, we find that a visible energy of $15$ MeV corresponds to a proton recoil of $\sim 21$ MeV. Using~\cref{eq:nuN1,eq:nuN2} for a proton instead of a xenon target, we find that, for $M_N=0.25$ GeV, the minimum neutrino energy to produce $N$ is $E_\text{th}^\text{p}\simeq 0.28$ GeV, and the corresponding recoil energy is $T_0^{\text{p}}\simeq 26$ MeV, which is already above the JUNO threshold recoil energy of $\sim 21$ MeV. 
This implies that for $M_N=0.25$ GeV (which is the heavy neutral lepton mass that we will later show to correspond to a viable explanation of the LZ event), neutrino energies across practically the entire atmospheric scale are large enough to induce an observable signal in JUNO through $\nu \,{\rm p} \to N \,{\rm p}$ scattering process.

This statement holds in general and is not strongly dependent on the mass of $N$. For example, let us take $M_N=1$ GeV, which is within the range for which the authors of Ref.~\cite{Jeesun:2026vzo} claim to have found a viable solution to the LZ anomaly. Using again \cref{eq:nuN1,eq:nuN2}, we find that the minimum neutrino energy for producing $N$ of such mass in LZ is $E_\text{th}^\text{Xe}\approx 1$ GeV, with a recoil of $T_0^{\text{Xe}}\approx \mathcal{O}(\text{MeV})$. We also find that the observed recoil energy at $\sim 250$ keV can be reached for atmospheric neutrino energies above $\approx 2.2$ GeV. As far as the scattering off protons at JUNO is concerned, for $M_N=1$ GeV, we obtain $E_\text{th}^\text{p}\simeq 1.5$ GeV and $T_0^{\rm p}\simeq 0.25$ GeV, which is already well above JUNO's $21$ MeV proton recoil threshold. Hence, also for the $M_N=1$ GeV benchmark point, we find that comparable ranges of the atmospheric neutrino spectrum are relevant for LZ and JUNO.

Therefore, differences in the flux and recoil energy ranges clearly cannot compensate for the five orders of magnitude larger exposure in JUNO. Taking JUNO as a representative example, we conclude that if $\nu\to N$ scattering at LZ were responsible for the observed event, many more events would be expected in neutrino detectors, yet such events have not been observed. This argument does not apply exclusively to JUNO~\cite{JUNO:2015zny}, which has only recently started taking data, but also to an array of previously and currently operating liquid scintillator detectors, such as KamLAND~\cite{KamLAND:2011bnd}, SNO+~\cite{SNO:2021xpa,SNO:2025chx}, and Borexino~\cite{Borexino:2008gab}, as well as Cherenkov detectors, such as Super-Kamiokande~\cite{Super-Kamiokande:2002weg,Super-Kamiokande:2023ahc} and IceCube~\cite{IceCube:2016zyt}, all of which have larger exposures than LZ.

In the remainder of this work, we present a scenario that not only mitigates but completely eliminates this seemingly bulletproof argument that a neutrino-induced LZ event would necessarily be accompanied by a large number of neutral-current events in neutrino detectors.

\section{The Minimal Viable Neutrino-Induced Realization}
\label{sec:N1-N2}
\noindent
Let us now consider the scattering of an incident particle $N_1$ of mass $M_1$ and energy $E$ off a nucleus of mass $m_A$.
The threshold energy for the process $N_1 A\to N_2 A$, in which $N_2$ with mass $M_2$ is produced, is given by
\begin{align}
    E_{\rm th}^A =M_2+\frac{M_2^2-M_1^2}{2m_A} \, ,
    \label{eq:Eth}
\end{align}
while the corresponding nuclear recoil energy reads
\begin{align}
    T^A_0= \frac{M_2^2-M_1^2}{2(m_A+M_2)} \, .
    \label{eq:Tstar}
\end{align}
One can also rewrite \cref{eq:Tstar} as
\begin{align}
    M_2=T_0^{\rm Xe}+\sqrt{M_1^2+2m_{\rm Xe}T_0^{\rm Xe}+\left(T_0^{\rm Xe}\right)^2} \, .
    \label{eq:M2needed}
\end{align}
Away from threshold $(E > E_{\rm th}^{A})$, the allowed nuclear recoil interval is $[T_A^-,T_A^+]$, where
\begin{align}
    T_A^{\pm} =\frac{X\pm\sqrt{\lambda_1\lambda_2}}{4m_As} \, .
    \label{eq:endpoints}
\end{align}
Here, $\lambda(x,y,z)=x^2+y^2+z^2-2xy-2xz-2yz$, and we denote $\lambda_i \equiv \lambda(s,M_i^2,m_A^2)$ with $s=M_1^2+m_A^2+2m_AE$. Further, $X=(s+M_1^2-m_A^2)(s+M_2^2-m_A^2) -2s(M_1^2+M_2^2)$. 

\begin{figure}[t!]
    \centering
    \includegraphics[width=0.9\linewidth]{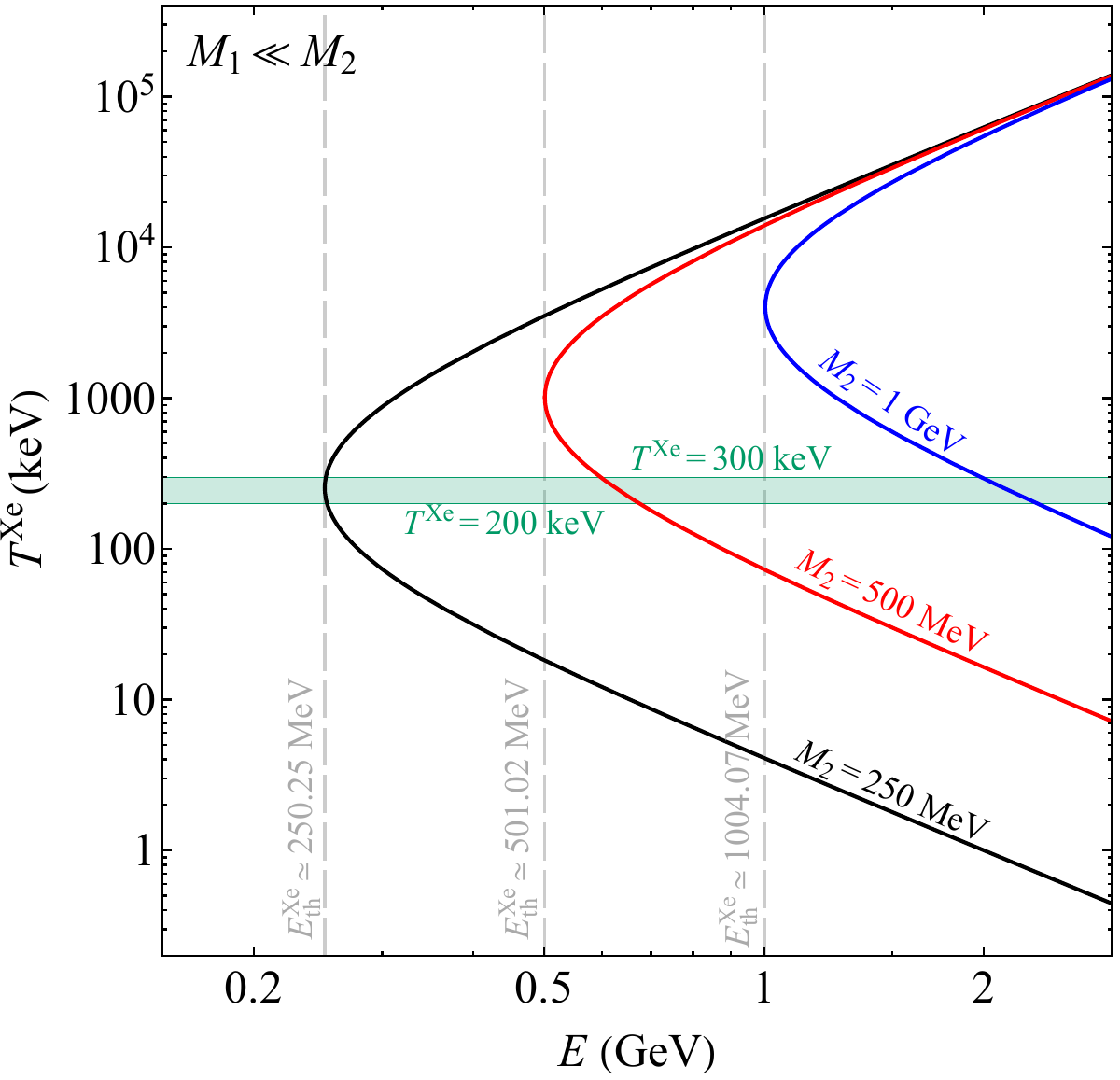}
    \caption{Xenon recoil energies arising from $N_1\,{\rm Xe}\to N_2 \, {\rm Xe}$ scattering process, shown for several values of $M_2$, are presented as a function of incoming energy of $N_1$. The horizontal green band marks the $200-300$~keV recoil region that contains $\simeq 250$ keV value reported by LZ. For each considered value of $M_2$, the dashed vertical lines indicate the corresponding threshold energies, $E_{\rm th}^\text{Xe}$.}
    \label{fig:recoil_range}
\end{figure}

In \cref{fig:recoil_range}, for $N_1 \,\text{Xe}\to N_2\, \text{Xe}$ we present both ``lower'' (the part of the parabola below its vertex)  and ``upper'' (above the vertex) nuclear recoil branches denoting allowed $[T_A^-,T_A^+]$ range as a function of incoming $N_1$ energy $E$. This is shown for several values of $M_2$; we assume $M_1$ mass to be much smaller than $M_2$. At the threshold energy, which for the case with $M_2=250$ MeV (black parabola) reads $E_\text{th}^\text{Xe}=250.25$ MeV, the only possible recoil energy is $T^A_0\simeq 250$ keV, while for larger energies the nuclear recoil energy range opens up.

\begin{figure}[t!]
\vspace{-0.05cm}
    \centering
    \includegraphics[width=0.9\linewidth]{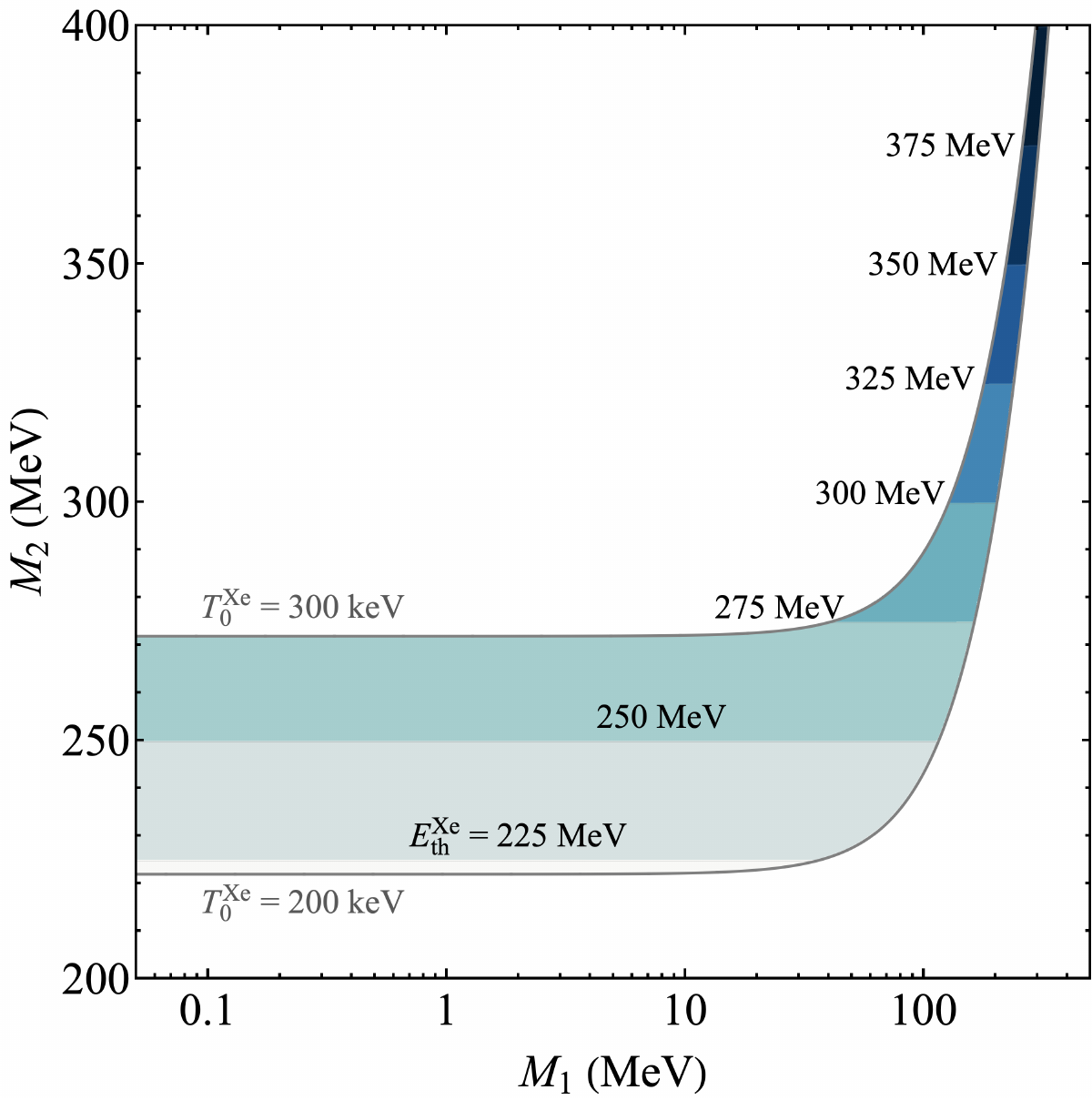}
    \caption{Region in the $(M_1,\,M_2)$ parameter space for which $N_1\,{\rm Xe}\to N_2 \, {\rm Xe}$ scattering at the threshold $N_1$ energy is achieved, limited to cases for which the xenon recoil is in the $(200-300)$~keV range. For the $(M_1,\,M_2)$ values in the shaded region, scattering off lighter targets is forbidden due to the larger $N_1$ energy required to surpass the threshold for scattering off such targets.}
\label{fig:masses}
\end{figure}

Let us focus on the vertices of the parabolas shown in \cref{fig:recoil_range}. 
For these points, we are right at the threshold, where a single energy $(E=E_\text{th}^\text{Xe})$ of incoming $N_1$ induces a nuclear recoil $T_0^{\rm Xe}$.   
From \cref{eq:Eth}, we observe that $E_\text{th}^A$ increases for lighter target nuclei.
Namely, for fixed $M_1$ and $M_2$, one finds
\begin{align}
    E_{\rm th}^{\rm Xe} <E_{\rm th}^{\rm O} <E_{\rm th}^{\rm C} <E_{\rm th}^{\rm p}\,.
    \label{eq:ordering}
\end{align}
For instance, the separation between the xenon and oxygen threshold energies is
\begin{equation}
    E_{\rm th}^{\rm O} - E_{\rm th}^{\rm Xe}
    =\frac{M_2^2-M_1^2}{2} \left( \frac{1}{m_{\rm O}}-\frac{1}{m_{\rm Xe}} \right) \, .
    \label{eq:window}
\end{equation}
For $M_1\simeq 0$ GeV, $M_2=0.25$ GeV, and given $m_{\rm Xe}\simeq123$~GeV and $m_{\rm O}=14.9$~GeV, $E_{\rm th}^{\rm O}-E_{\rm th}^{\rm Xe}\simeq1.8$~MeV. Thus, there is a narrow energy window in which $N_1 \,\text{Xe}\to N_2 \,\text{Xe}$ is allowed while $N_1 \,\text{O}\to N_2 \,\text{O}$ is kinematically forbidden. The same holds for scattering off any nuclei lighter than xenon. In particular, the energy ranges in which scattering on xenon is kinematically allowed while scattering on carbon and protons is forbidden are approximately $2.5$ MeV and $33$ MeV wide, respectively.

In \cref{fig:masses}, for the $N_1 \,\text{Xe}\to N_2 \,\text{Xe}$ process, we present the region in the $M_1$-$M_2$ parameter space in which scattering at the threshold is realized. Since LZ reported a $\sim 250$ keV recoil, we consider a threshold recoil range between $T_0^\text{Xe}=200$ keV and $T_0^\text{Xe}=300$ keV, with the color code indicating the corresponding $E_\text{th}^\text{Xe}$ values. Throughout the shaded region, scattering on lighter elements such as oxygen and carbon is kinematically forbidden.

\begin{table}[b!]
\centering
\small
\renewcommand{\arraystretch}{1.2}
\begin{tabular}{|l|c|c|}
\hline
Hadronic final state
& $E_{A}^{\star}~({\rm MeV})$
& $E_{\mathrm{th}}^{A^\star}~({\rm MeV})$ \\
\hline
\hline
$^{40}{\rm Ar}\to{}^{40}{\rm Ar}^{*}(1.46~{\rm MeV}\,\gamma)$
& 1.461 & 252.31 \\
\hline
$^{131}{\rm Xe}\to n+{}^{130}{\rm Xe}$
& 6.604 & 256.874 \\
\hline
$^{12}{\rm C}\to{}^{12}{\rm C}^{*}(4.44~{\rm MeV}\,\gamma)$
& 4.440 & 257.336 \\
\hline
$^{16}{\rm O}\to{}^{16}{\rm O}^{*}(6.05~{\rm MeV}\,e^+e^-)$
& 6.049 & 258.250 \\
\hline
$^{16}{\rm O}\to{}^{16}{\rm O}^{*}(6.13~{\rm MeV}\,\gamma)$
& 6.130 & 258.332 \\
\hline
$^{132}{\rm Xe}\to n+{}^{131}{\rm Xe}$
& 8.937 & 259.210 \\
\hline
$^{40}{\rm Ar}\to n+{}^{39}{\rm Ar}$
& 9.869 & 260.78 \\
\hline
$^{40}{\rm Ar}\to p+{}^{39}{\rm Cl}$
& 12.529 & 263.45 \\
\hline
$^{16}{\rm O}\to p+{}^{15}{\rm N}$
& 12.127 & 264.434 \\
\hline
$^{16}{\rm O}\to n+{}^{15}{\rm O}$
& 15.664 & 268.033 \\
\hline
$^{12}{\rm C}\to p+{}^{11}{\rm B}$
& 15.957 & 269.122 \\
\hline
$^{16}{\rm O}\to p+{}^{15}{\rm N}^{*}(6.32~{\rm MeV}\,\gamma)$
& 18.451 & 270.870 \\
\hline
$^{12}{\rm C}\to n+{}^{11}{\rm C}$
& 18.721 & 271.952 \\
\hline
\end{tabular}
\caption{Representative excitation and breakup channels for
$M_1\simeq0$~MeV and $M_2=250$~MeV. Here $E_{A}^{\star}$ is the
minimum energy above the nuclear ground state required to access the
channel, and $E_{\mathrm{th}}^{A^\star}$ is the corresponding incident
$N_1$ threshold.}
\label{tab:thresholds}
\end{table}

\begin{figure}[t!]
\vspace{0.05cm}
    \centering
    \includegraphics[width=1\linewidth]{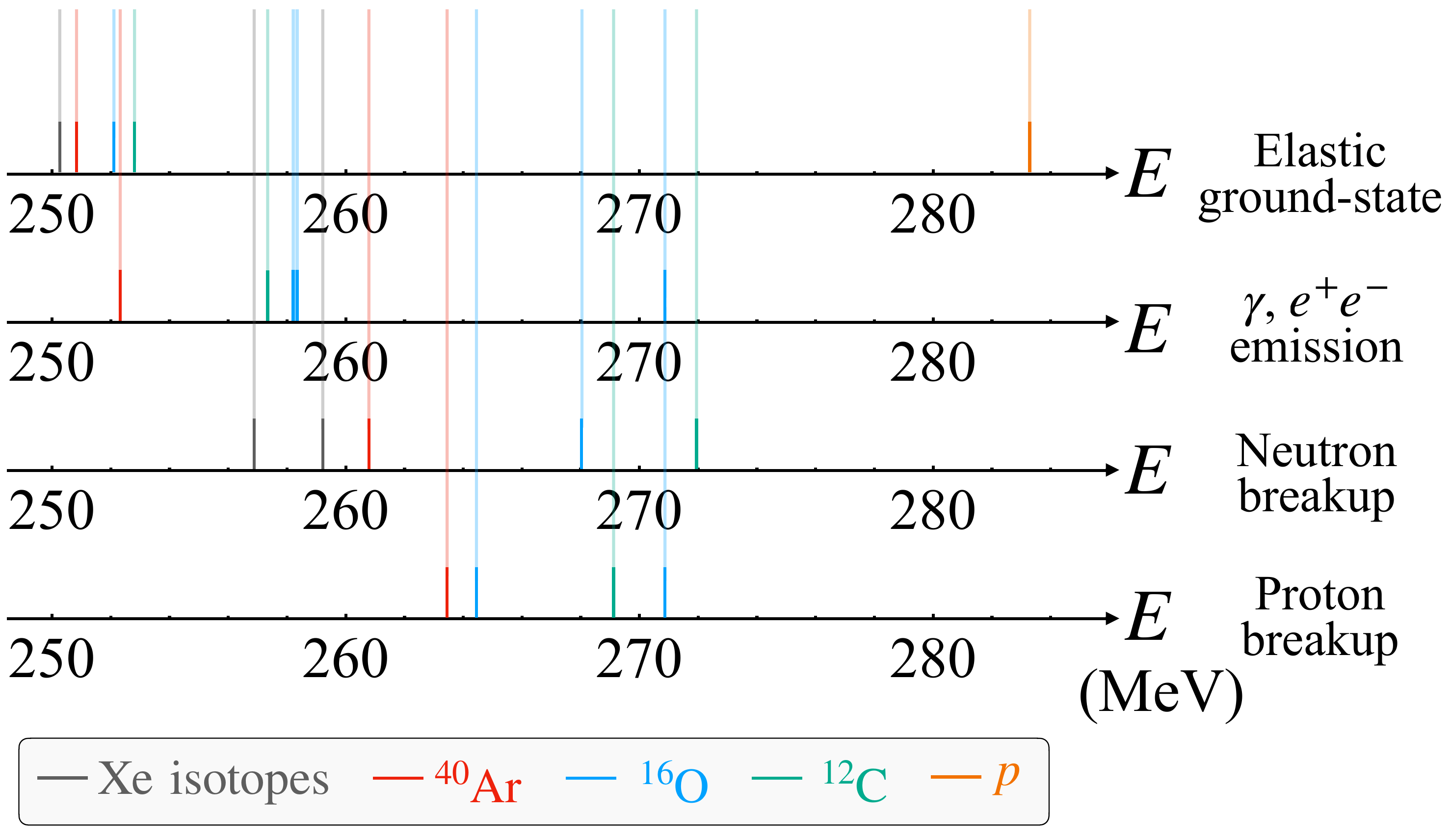}
    \caption{Energy thresholds for $N_1 A$ scattering into $N_2$ and various nuclear final states for the benchmark values of $M_1\simeq0$~MeV and $M_2=250$~MeV. The vertical lines denote energy thresholds for elastic ground-state scattering, nuclear excitation followed by $\gamma$ or $e^+e^-$ emission, and neutron and proton breakup for the relevant target nuclei.}
    \label{fig:realthreshold}
\end{figure}

\cref{eq:Eth} is applicable to the case in which the nucleus remains in its ground state following the scattering process. At higher incoming $N_1$ energies, however, there may be sufficient energy for the nucleus to transition to an excited state or for processes involving neutron or proton breakup, in which a nucleon is knocked out of the nucleus. The threshold energies for such processes, summarized in \cref{fig:realthreshold} for $M_1\simeq0$ MeV and $M_2=250$~MeV benchmark point, are necessarily higher than those for the processes in which the nucleus remains in its ground state (also shown in the figure for comparison). We computed these values using the excitation energies, $E_A^*$, of the relevant nuclear excited states tabulated in Refs.~\cite{Ajzenberg-Selove:1991rsl,Tilley:1993zz,Kelley:2017qgh,Chen:2017ngq,Wang:2021xhn}, and \cref{eq:Eth} with the replacement $M_2\to M_2+E_A^*$. In \cref{tab:thresholds}, we list representative excitation and breakup channels together with the corresponding nuclear excitation energies and incident $N_1$ thresholds for the benchmark point $M_1\simeq0$~MeV and $M_2=250$~MeV.

Hence, if one considers the energy window in which scattering in neutrino experiments with the nucleus remaining in the ground state is forbidden, this automatically means that $N_1$ does also not have enough energy to excite the nucleus or eject a nucleon. Thus, in order to have interactions in xenon while avoiding interactions in oxygen for our benchmark point, $M_1\simeq 0$ MeV and $M_2=250$ MeV, it is sufficient to ensure that the $N_1$ energy range does not exceed $E=E_{\rm th}^{\rm Xe}+1.8$ MeV, requiring $N_1$ to have a very narrow, nearly monoenergetic energy spectrum.

This is precisely the idea behind our neutrino-based explanation of the LZ event: the interaction is kinematically allowed in a xenon detector while remaining kinematically inaccessible in large neutrino detectors. In \cref{sec:narrow}, we will demonstrate how to achieve a nearly monoenergetic $N_1$ spectrum, as otherwise, when considering the full spectrum, the shortcomings discussed in \cref{sec:nu-N+kinematics} apply.

Before doing so, note that our discussion above was chiefly based on oxygen, carbon, and hydrogen, which are typical targets in large-scale neutrino experiments. We have, however, not discussed argon, which is heavier than all of these targets and therefore leads to a smaller window, $E_{\rm th}^{\rm Ar}-E_{\rm th}^{\rm Xe}\simeq 0.6$~MeV, for our representative benchmark point, $M_1\simeq 0$ MeV and $M_2=250$ MeV.
If the $N_1$ spectrum extends above the argon threshold while remaining below the oxygen threshold, with a fraction of the generated $N_1$ flux in the range $E_{\rm th}^{\rm Xe}+0.6~{\rm MeV} \lesssim E \lesssim E_{\rm th}^{\rm Xe}+1.8~{\rm MeV},$ the upcoming DUNE experiment may be able to probe the parameter space that explains the LZ event using its atmospheric neutrino data \cite{DUNE:2020ypp,DUNE:2026yly}, including sub-GeV samples \cite{Kelly:2019itm}. 
We also note the DEAP-3600 liquid-argon dark-matter experiment~\cite{DEAP:2019yzn,DEAPCollaboration:2021raj}, which has already collected data. Near threshold, however, the argon recoil energies can lie outside the range used in existing WIMP searches; namely, for our benchmark, scattering starts at $T^{\rm Ar}_0\simeq 830$~keV, whereas the DEAP-3600 WIMP search is sensitive to roughly $50-200$~keV nuclear recoils~\cite{DEAP:2026orr,Baer:2026yrt}. In addition, its tonne-scale target implies a much smaller overall event yield than in future larger argon detectors, ranging from the multi-tonne DarkSide-20k experiment~\cite{DarkSide-20k:2017zyg} to kilotonne-scale neutrino detectors such as aforementioned DUNE~\cite{DUNE:2020ypp}. Overall, the most robust future tests of the narrow $N_1$ energy spectrum may come from argon-based experiments.

\section{Generating a Nearly Monoenergetic $\mathbf{N_1}$ Spectrum}
\label{sec:narrow}
\noindent
In what follows, we explore scenarios for producing a nearly monoenergetic $N_1$ spectrum from the broad spectrum of atmospheric neutrinos. The idea is to confine the $N_1$ spectrum to the energy range in which scattering $N_1\,\text{Xe}\to N_2\,\text{Xe}$ can occur, while scattering off lighter targets such as carbon and oxygen is kinematically forbidden.
We start by considering the conversion of atmospheric neutrinos to $N_1$ through a matter-induced resonance (\cref{subsec:1}), followed by parametric resonance in a DM background (\cref{subsec:2}).

\subsection{Conversion Through a Matter-Induced Resonance}
\label{subsec:1}
\noindent
There is a resonant effect for multi-GeV atmospheric neutrinos in the Earth that enhances flavor conversion~\cite{Akhmedov:1998xq}. In our case, where atmospheric neutrinos convert to $N_1$, which is assumed to be much more massive than active neutrinos, the resonance could only occur for a much larger matter potential (see, for instance, Ref.~\cite{Brdar:2023cms}, where such matter potential was studied in the context of the gallium anomaly). 
A dark sector or an ultralight DM background interacting with $N_1$ can generate such a potential, dubbed $V$~\cite{Bramante:2011uu,Kopp:2014fha,Capozzi:2017auw,Brdar:2023cms,Brdar:2025azm}.
Working in a two-flavor approximation, namely, with one active neutrino and the $N_1$ state, the effective mixing angle in matter is defined as \cite{Wolfenstein:1977ue,Mikheyev:1985zog,Mikheev:1986wj}
\begin{align}
    \sin^2 2\theta_m =\frac{w^2}{(E-E_R)^2+w^2} \,,
    \label{eq:dark_msw_mixing}
\end{align}
where $E_R = \Delta m^2\cos2\theta/(2V)$ is the resonance energy, $\Delta m^2\equiv M_1^2-m_\nu^2\simeq M_1^2$, $\theta$ is the vacuum mixing angle for the $\nu$--$N_1$ system and  $w \equiv E_R \tan 2\theta$. The resonance therefore facilitates flavor conversion, which is most pronounced around $E_R$.
 
The flavor conversion probability reads
\begin{align}
 P_{\nu N_1}^{\text{MSW}}(E,L) = \sin^2 2\theta_m\, \sin^2\left(\frac{\Delta \widetilde{m}^2 L}{4E}\right) \; ,
\end{align}
with
\begin{align}
 \Delta \widetilde{m}^2= \sqrt{(\Delta m^2\cos2\theta-2E V)^2+(\Delta m^2\sin2\theta)^2} \, .
\end{align}
The sign of $V$ determines whether the resonance appears in the neutrino or antineutrino sector. When the oscillation phase is averaged, the transition probability can be expressed as
\begin{equation}
  \langle P_{\nu N_1}^{\text{MSW}}(E) \rangle= \frac{1}{2} \frac{w^2}{(E-E_R)^2+w^2} \, .
 \label{eq:msw_lorentzian}
\end{equation}


\begin{figure}[t!]
\centering
\includegraphics[width=0.9\linewidth]{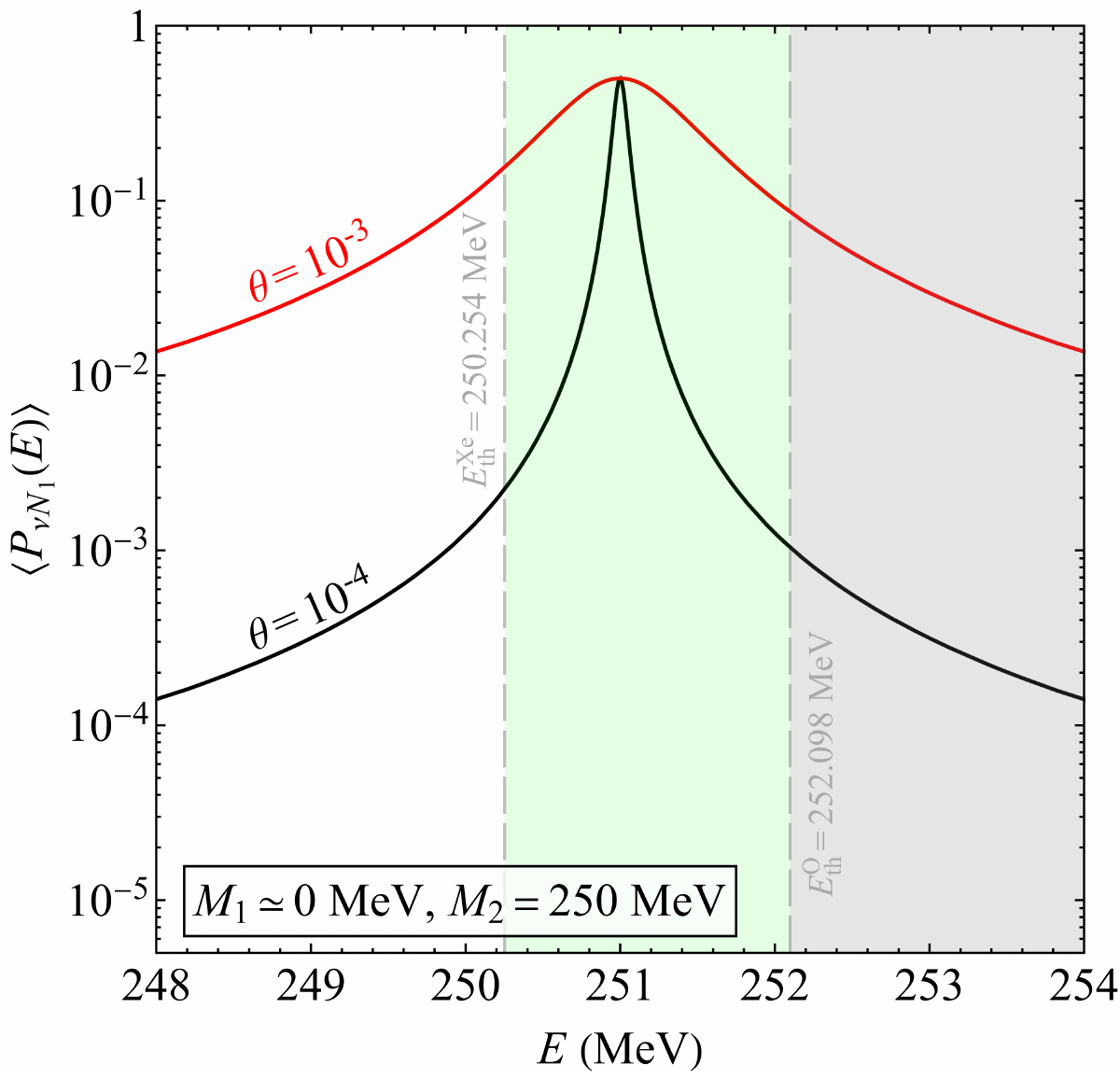}
\caption{Phase-averaged $\nu\to N_1$ conversion probability for two values of the vacuum mixing angle. The green shaded region between the xenon and oxygen threshold energies indicates the phenomenologically viable energy window for the benchmark $M_1\simeq0$ MeV, $M_2=250$~MeV. Smaller $\theta$ narrows the resonance and suppresses the high-energy tail.}
\label{fig:msw_spectrum}
\end{figure}


\cref{fig:msw_spectrum} shows this averaged probability as a function of neutrino energy, evaluated for two values of the vacuum mixing angle $\theta$. As can be observed from the figure, the resonance width strongly depends on the vacuum mixing angle; it decreases for smaller values of $\theta$, see also \cref{eq:dark_msw_mixing}.
While, for small $\theta$, the produced $N_1$ flux is confined more tightly to the desired energy window below the oxygen threshold,  the flavor conversion is overall reduced (compare areas under red and black line in \cref{fig:msw_spectrum}). Such a decrease in the produced $N_1$ flux would require a larger interaction rate for the $N_1\,\text{Xe}\to N_2\,\text{Xe}$ process in order to explain the observed LZ event.

Furthermore, at sufficiently small $\theta$ and for $m_\nu < M_1$, the oscillation length at resonance
\begin{align}
 L_{\rm res}
 = &\, \frac{4\pi E_R}{\Delta m^2 \sin2\theta} \nonumber\\
 \simeq &\, 310\,{\rm km}\,
 \left(\frac{E_R}{250~{\rm MeV}}\right)
 \left(\frac{10~{\rm eV}}{M_1}\right)^2
 \left(\frac{10^{-5}}{\theta}\right) \, ,
\end{align}
becomes comparable to the Earth radius, implying that there is not enough path length to efficiently convert an active neutrino into $N_1$, despite meeting the resonance condition. This issue can be evaded by increasing $M_1$; for example, for $\theta=10^{-6}$, requiring the resonant oscillation length to satisfy $L_{\rm res}\lesssim 1000\,{\rm km}$, a representative atmospheric-neutrino baseline, gives $M_1\gtrsim\mathcal{O}(10)$~eV. However, increasing $M_1$ requires a larger matter potential in order to satisfy the resonant condition, $E=E_R$, for atmospheric neutrinos.

Still, the main problem with producing $N_1$ through a matter-induced resonance remains the resonance width. Note that for neutrino energies much larger than $E_R$, $\langle P_{\nu N_1}^{\rm MSW}(E)\rangle \sim (\omega/E)^2$, leaving a not-too-strongly-falling high-energy tail that produces an $N_1$ flux above the oxygen, carbon, and proton threshold energies. This, in turn, potentially induces a number of neutral-current $N_1 A\to N_2 A$ events at large-scale neutrino experiments.

\begin{figure}[t!]
\centering
\includegraphics[width=0.9\linewidth]{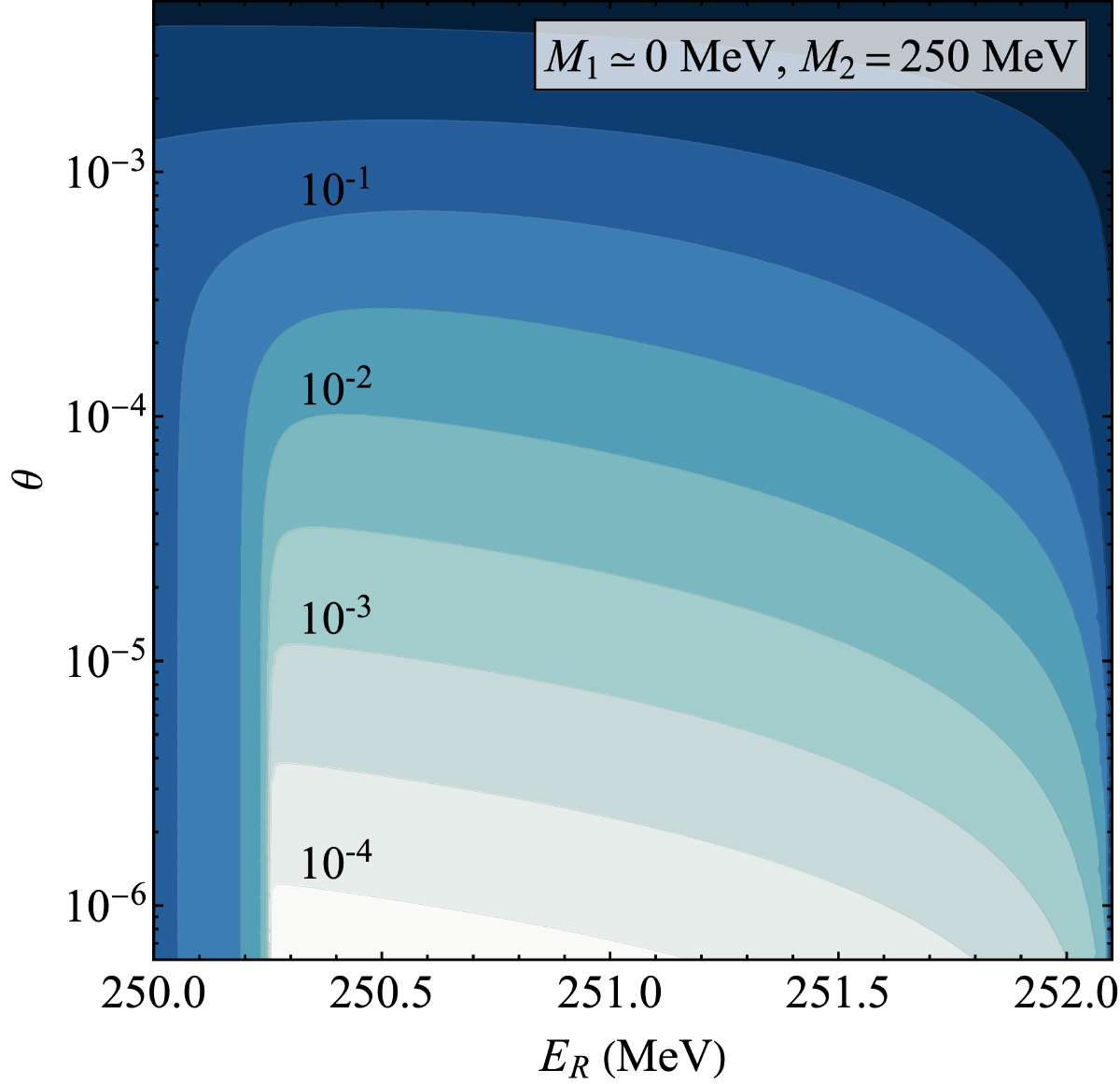}
\caption{Ratio of the generated $N_1$ flux above the oxygen threshold, $E_\text{th}^\text{O}$, to that between the xenon and oxygen thresholds, $E_\text{th}^\text{Xe}<E<E_\text{th}^\text{O}$. This quantity, defined in \cref{eq:msw_tail_ratio} as $R_{\Phi}$, is shown as a function of the resonance energy $E_R$ and for several values of vacuum mixing angle $\theta$. Smaller mixing angles lead to a narrower resonance (see also \cref{fig:msw_spectrum}) and therefore suppress the $N_1$ flux above $E_\text{th}^\text{O}$.}
\label{fig:msw_ratio}
\end{figure}

Let us quantify this high-energy tail: we define the ratio of the generated $N_1$ flux above the threshold energy for oxygen, $E_\text{th}^\text{O}\approx 252.1$ MeV, to the $N_1$ flux generated between the xenon and oxygen threshold energies, for the benchmark point $M_2= 250$ MeV and $M_1\ll M_2$. We therefore have
\begin{align}
    \mathcal R_{\Phi}(E_R,\theta)
    =
    \frac{\int_{E_\text{th}^\text{O}}^{3\,{\rm GeV}} dE\,\Phi_{N_1}(E)}
    {\int_{E_\text{th}^\text{Xe}}^{E_\text{th}^\text{O}} dE\, \Phi_{N_1}(E)\,}\,,
    \label{eq:msw_tail_ratio}
\end{align}
with the $N_1$ flux given by $\Phi_{N_1}(E) \equiv \Phi_{\nu} (E) \langle P_{\nu\to N_1}^{\rm MSW}(E)\rangle $. In this energy range, we model the atmospheric neutrino spectrum as $\Phi_\nu(E)\propto E^{-2.7}$ (for a detailed flux calculation, see
Ref.~\cite{Honda:2015fha}). $\mathcal R_{\Phi}$ directly measures how much $N_1$ flux is produced beyond the desired energy window relative to the flux available to generate the LZ signal.
It is shown in \cref{fig:msw_ratio} as a function of $E_R$ and for several values of $\theta$. From the figure, one can observe that suppressing the $N_1$ population above the oxygen threshold ($E_\text{th}^\text{O}\approx 252.1$ MeV) requires increasingly smaller $\theta$. For such values of vacuum mixing angle, as discussed above, the produced $N_1$ flux is strongly suppressed, and a larger matter potential is also needed in order to avoid too large an oscillation length at resonance. We therefore find conversion through a matter-induced resonance to be inadequate for producing sufficiently large $N_1$ flux below threshold energies for scattering off targets used in large-scale neutrino experiments.


\subsection{Parametric Resonance}
\label{subsec:2}
\noindent
Let us now turn to a different realization which employs a scalar field $\phi$ constituting the DM. The relevant part of the Lagrangian is~\cite{Chun:2021ief,Losada:2022uvr,Brdar:2023cms}
\begin{align}
 \mathcal L
 \supset & \,
 \frac12(\partial_\mu\phi)(\partial^\mu\phi)
 -\frac12m_\phi^2\phi^2 \nonumber\\
 & \qquad -\left[
   \frac12 M_1 \bar{N}^c_1 N_1+ g \bar{N}_1\phi\,\nu
   +{\rm h.c.}
  \right] \, ,
 \label{eq:scalar_lagrangian}
\end{align}
where $N_1$ is assumed to be a Majorana fermion, $m_\phi$ is the mass of the $\phi$ field, and $g$ is the coupling between $\phi$, $N_1$, and the active neutrinos.

In the classical background, and in two-flavor approximation, the neutral fermion mass matrix reads
\begin{align}
 \mathcal M(\phi)=
 \begin{pmatrix}
  0 & g\phi\\
  g\phi & M_1
 \end{pmatrix} \, .
 \label{eq:neutral}
\end{align}
From this matrix, one can infer that $\nu$--$N_1$ mixing remains nonvanishing even when the vacuum mixing angle is zero. Therefore, the presence of ultralight DM background field and the interaction term in \cref{eq:scalar_lagrangian} facilitates flavor transitions between active neutrinos and $N_1$. This effect, dubbed parametric resonance, was previously adopted in Refs. \cite{Chun:2021ief,Losada:2022uvr,Brdar:2023cms}.

The classical field $\phi$ is described as 
\begin{equation}
 \phi(t)=\frac{\sqrt{2\rho_\phi}}{m_\phi} \cos(m_\phi t+\varphi)\,,
\end{equation}
where $\rho_\phi$ is the local DM energy density that is approximately $0.4$ GeV $\text{cm}^{-3}$ and $\varphi$ is oscillation phase.

The resonance for $\nu$--$N_1$ transition occurs when the condition
\begin{equation}
 \frac{M_1^2}{2E_R}=m_\phi\, ,
 \label{eq:scalar_resonance}
\end{equation}
is satisfied.  Here, as in \cref{subsec:1}, $E_R$ denotes the neutrino energy required to achieve the resonance.

A single coherent mode produces oscillatory conversion, giving \cite{Losada:2022uvr}
\begin{equation}
 P_{\nu\to N_1}^{\rm coh} = \frac{\Omega^2}{\delta^2+\Omega^2} \sin^2\left[ \frac{L}{2}\sqrt{\delta^2+\Omega^2} \right] \, ,
\end{equation}
where
\begin{equation}
 \delta \equiv \frac{M_1^2}{2E}-m_\phi \, , \qquad
 \Omega \equiv \frac{gM_1}{2E} \frac{\sqrt{2\rho_\phi}}{m_\phi} \, .
\end{equation}

Previous studies of parametric neutrino conversion induced by ultralight DM have generally treated the scalar background as a coherent oscillating field. Here we account for the finite velocity dispersion of the DM field, which induces Doppler broadening and a finite coherence length along the neutrino trajectory. 

We model the scalar background as an ensemble with a nonrelativistic velocity distribution. Along a neutrino trajectory, a scalar mode is sampled with the effective frequency
\begin{equation}
    \omega_{\rm path}
    =
    m_\phi-\boldsymbol{k}_\phi\cdot\hat{\boldsymbol n}
    \simeq
    m_\phi(1-u) \,,
\end{equation}
where $\boldsymbol{k}_\phi$ is the scalar field's wave vector with magnitude $m_\phi v_\phi$, $\hat{\boldsymbol n}$ is the unit vector in the direction of neutrino propagation, and $ u\equiv\boldsymbol v_\phi\cdot\hat{\boldsymbol n}$ is the velocity component along the neutrino trajectory.

We take the scalar velocities to follow a truncated Maxwellian line-of-sight velocity distribution
\begin{equation}
    f_\parallel(u)=
    \frac{e^{-u^2/v_0^2}-e^{-v_{\rm esc}^2/v_0^2}}
    {\sqrt{\pi}\,v_0\,\mathcal N_{\rm esc}}
    \Theta(v_{\rm esc}-|u|)\,,
\end{equation}
with
\begin{equation}
    \mathcal N_{\rm esc}
    =
    {\rm erf}(z)-\frac{2z}{\sqrt{\pi}}e^{-z^2} \, ,
    \qquad
    z=\frac{v_{\rm esc}}{v_0} \,.
\end{equation}
Here, $v_0=220$ km/s is the speed of the locally virialized DM halo~\cite{Koposov:2009hn,Strigari:2012acq}, while
$v_{\text{esc}}=544$ km/s is the escape velocity~\cite{Baxter:2021pqo}.

The scalar two-point correlation function can be expressed as
\begin{align}
 C_\phi(\tau)=\left\langle \phi(t)\phi(t+\tau)\right\rangle=
 \frac{\rho_\phi}{m_\phi^2} \!
 \int \! du\,f_\parallel(u)
 \cos[m_\phi(1-u)\tau]\,,
 \label{eq:correlator}
\end{align}
where $\tau$ is the time separation between two points on the neutrino's trajectory.

Using \cref{eq:correlator}, we find the $\nu$--$N_1$ transition strength, accounting for both resonant and off-resonant transitions
\begin{align}
\begin{aligned}
\mathcal T(E,L)=\Omega^2
\int_{-v_{\rm esc}}^{v_{\rm esc}} &\, du \,f_\parallel(u)
\bigg[
\frac{\sin^2[(\delta+m_\phi u)L/2]}
     {(\delta+m_\phi u)^2}\\
+&\frac{\sin^2[(\delta+2m_\phi-m_\phi u)L/2]}
      {(\delta+2m_\phi-m_\phi u)^2}
\bigg]\,.
\end{aligned}
\label{eq:T}
\end{align}
The transition probability then reads
\begin{align}
P_{\nu N_1}(E,L)\simeq
\frac12\left[1-e^{-2\mathcal T(E,L)}\right] \, .
\label{eq:P}
\end{align}

In obtaining \cref{eq:T,eq:P}, we require the conversion to remain weak within each coherence interval, namely $\Omega_R\ell_{\rm coh}\ll1$, while considering propagation over distances $L\gg\ell_{\rm coh}$. Here, $ \Omega_R = g\sqrt{2\rho_\phi} / M_1$ and $\ell_{\rm coh}=1/(m_\phi\sigma_u)$ is the characteristic coherence length of the scalar field, with $\sigma_u$ denoting the standard deviation of the DM velocity distribution.
We additionally require $L\ll\ell_{\rm dec}$, where $\ell_{\rm dec}(E)\simeq E/(M_1\Gamma_{N_1})$ is the characteristic decay length of $N_1$ in the laboratory frame, with
\begin{equation}
    \Gamma_{N_1}
    =
    \frac{g^2 M_1}{16\pi}
    \left(1-\frac{m_\phi^2}{M_1^2}\right)^2 \, .
\end{equation}

Despite the requirement of weak conversion within each coherence interval, the total conversion probability can nevertheless become large, approaching $1/2$ when $2\mathcal{T}(E,L)\gg1$. A full numerical calculation would consist of evolving the neutrino state through many realizations of the scalar field and averaging the resulting conversion probabilities. Assuming Gaussian field statistics with correlation $C_\phi(\tau)$, we find very good agreement with the semi-analytic approximation presented in \cref{eq:T} throughout the $(m_\phi,g)$ region of interest, including the off-resonant tails.

\begin{figure*}[t!]
    \centering
    \includegraphics[width=0.317\linewidth]{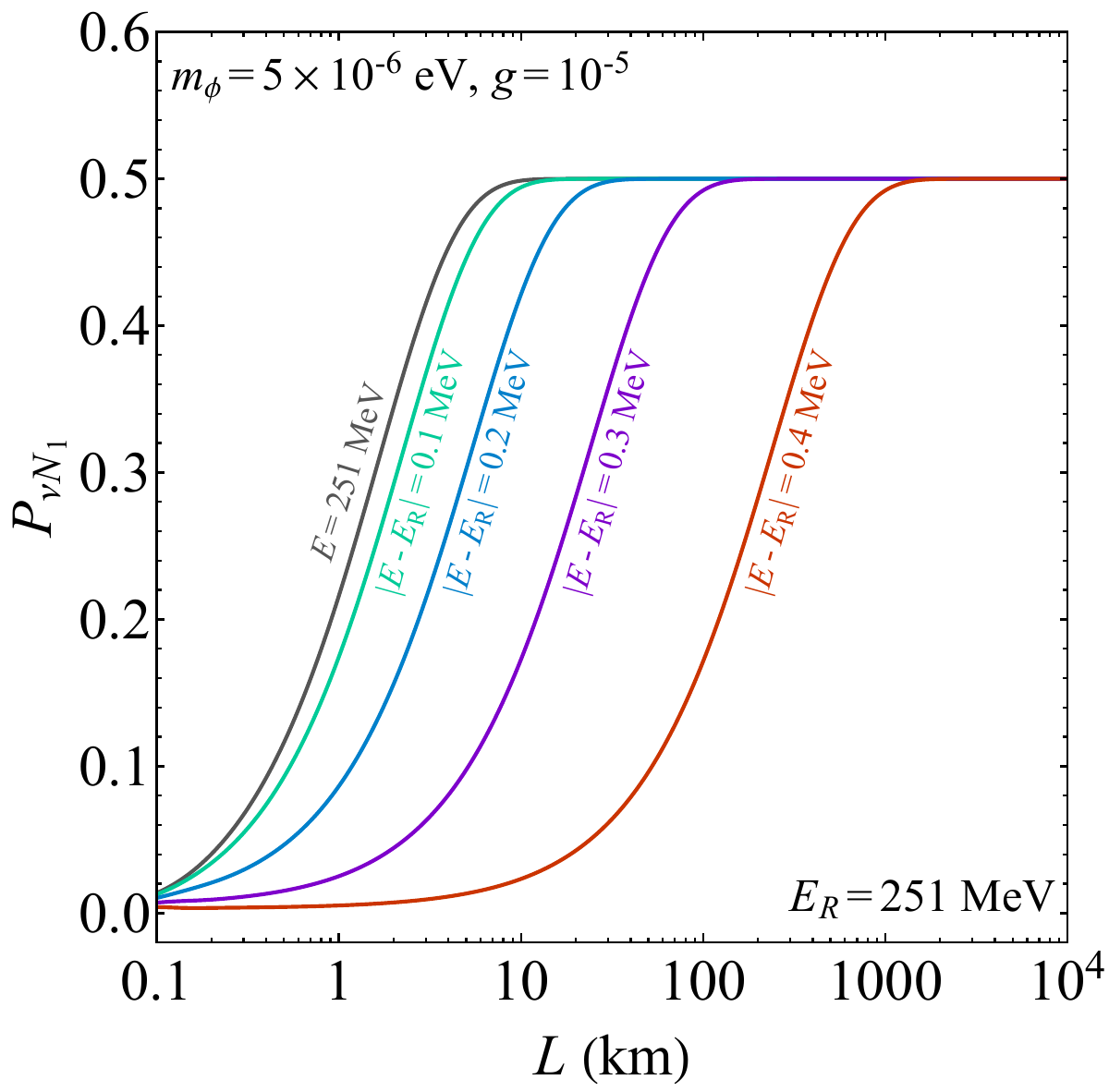}
    \includegraphics[width=0.335\linewidth]{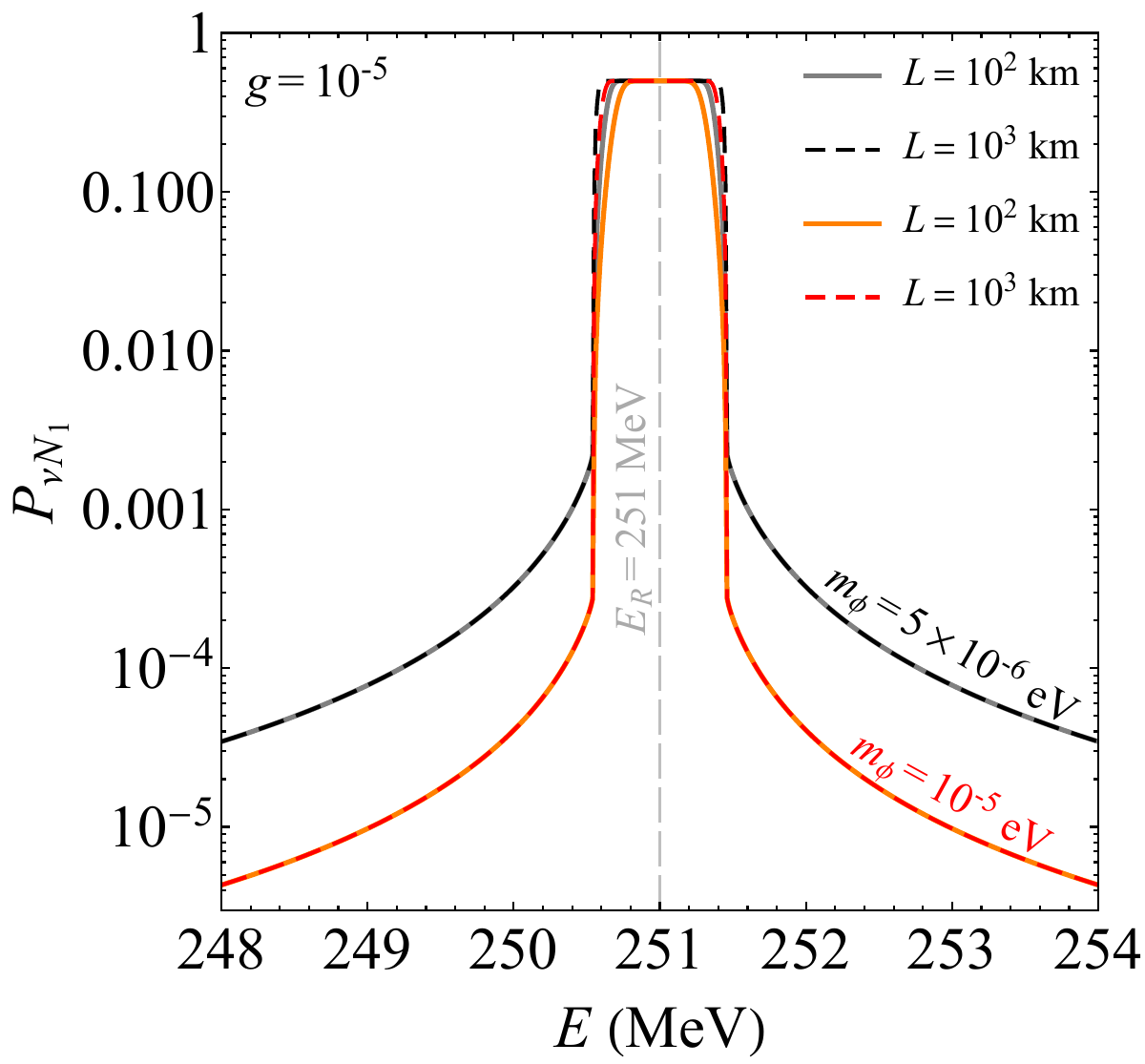}
    \includegraphics[width=0.335\linewidth]{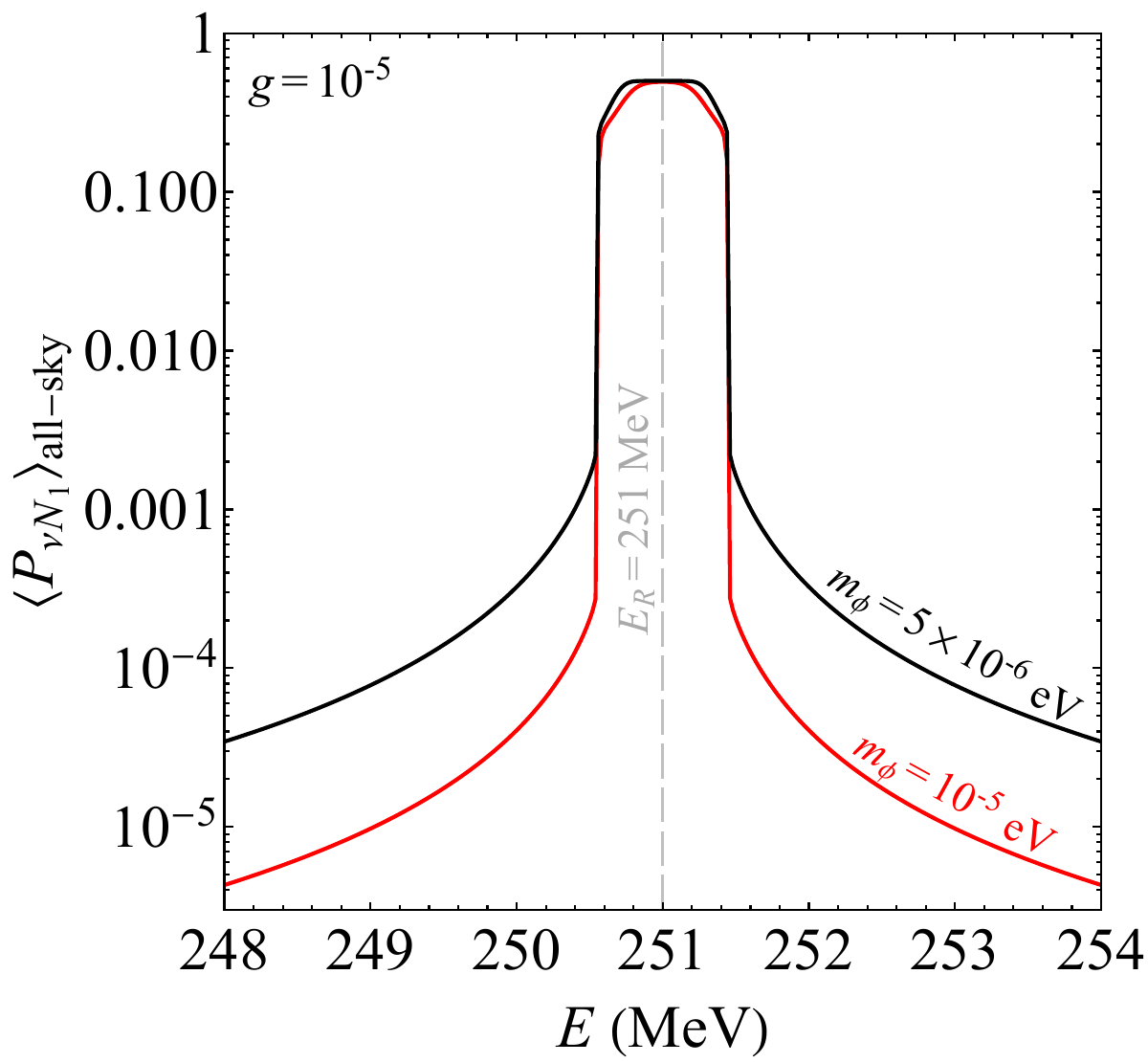}
    \caption{$\nu\to N_1$ conversion probability via parametric resonance in a DM background. \emph{Left}: conversion probability as a function of baseline for the indicated energy offsets from resonance, with $m_\phi=5\times10^{-6}$ eV and $g=10^{-5}$. \emph{Middle}: conversion probability as a function of energy for different baselines and two values of $m_\phi$, with the same $g$. \emph{Right}: all-sky-averaged conversion probability, obtained by accounting for the angular dependence of the atmospheric-neutrino baseline.}
\label{fig:parametric_conversion}
    \label{fig:resonance_prob}
\end{figure*}

In the left panel of \cref{fig:parametric_conversion}, we present the $\nu\to N_1$ conversion probability for $g=10^{-5}$ and $m_\phi=5\times10^{-6}$ eV as a function of baseline for the indicated energy offsets from resonance. At resonance, the probability approaches $1/2$ over only a few kilometers, whereas neutrinos away from resonance require longer baselines because fewer DM modes satisfy the resonance condition. In the middle panel, we show the conversion probability as a function of neutrino energy for different baselines.
Comparing with the matter-induced resonance scenario in \cref{subsec:1}, and in particular with \cref{fig:msw_spectrum}, we find less pronounced off-resonant tails, resulting in stronger suppression of the flux above the oxygen threshold. We also find that the conversion probability remains large over a wider region around $E_R$, owing to the Doppler broadening effect. This is particularly pronounced for $m_\phi=10^{-5}\,\mathrm{eV}$, for which the conversion probability remains approximately constant in a narrow region around $E=251\,\mathrm{MeV}$ and then drops by more than five orders of magnitude just $2\,\mathrm{MeV}$ away from resonance.

To obtain the expected $N_1$ spectrum from $\nu \to N_1$ transitions, we average the conversion probability over atmospheric neutrino trajectories. Approximating the atmospheric neutrino production region by a shell at altitude $h$, the path length for a neutrino arriving at a zenith angle $\theta$ reads
\begin{align}
    L(\cos \theta)=
    \sqrt{R_\oplus^2\cos^2 \theta+2R_\oplus h+h^2}
    -R_\oplus \cos \theta\,,
\end{align}
where $\cos \theta=1$ corresponds to vertically downgoing neutrinos. For an isotropic flux,
\begin{align}
    \langle P_{\nu N_1}(E) \rangle_{\rm all-sky}=
    \frac12\int_{-1}^{1}\!\! d\cos\theta\,
    P_{\nu N_1}[E,L(\cos \theta)]\,.
    \label{eq:sky_average}
\end{align}
We take $R_\oplus=6371\,{\rm km}$ and an atmospheric neutrino production height of $h=15\,{\rm km}$.

The all-sky-averaged conversion probability is shown in the right panel of \cref{fig:parametric_conversion}. As in the middle panel, where the conversion probability is shown for several baselines, we find that for the considered benchmark points, particularly for $m_\phi=10^{-5}\,\mathrm{eV}$, the flux is strongly confined to the narrow energy region below the oxygen threshold. Using the $\mathcal R_{\Phi}$ measure from \cref{eq:msw_tail_ratio}, we find $9.66\times10^{-5}$ for $m_\phi=10^{-5}$~eV and $6.83\times10^{-4}$ for $m_\phi=5\times10^{-6}$~eV, with $E_R=251$~MeV.

For a benchmark of $m_\phi=10^{-5}$~eV and $g= 10^{-5}$, we obtain an all-sky averaged conversion probability of $ \langle P_{\nu N_1}(E) \rangle_{\rm all-sky} \approx 0.35$ in the $1$~MeV window around the resonance.
In what follows, we use this value to determine the interaction strength necessary to produce a single event at LZ.

\section{Explaining the LZ Event via $\mathbf{N_1}$ Scattering}
\label{sec:N1-N2-xsec}
\noindent
Following the computation of $\langle P_{\nu N_1}\rangle_{\rm all-sky}$, we are now ready to compute the $N_1$ flux, denoted by $\Phi_1$, at the location of the LZ detector. We have
\begin{align}
\frac{d\Phi_1}{dE}= \langle P_{\nu N_1}\rangle_{\rm all-sky} \frac{d \Phi_\nu}{dE}\,,
\end{align}
where $d\Phi_\nu/dE$ is the total atmospheric neutrino flux, which at an energy of $E=250$ MeV reads $d\Phi_\nu/dE\simeq 14\,\mathrm{cm}^{-2}\,\mathrm{s}^{-1}\,\mathrm{GeV}^{-1}$\cite{Honda:2015fha,Zhuang:2021rsg}.
This yields
\begin{align}
    \frac{d\Phi_1}{dE} \simeq 5\times10^{-3} \, {\rm cm}^{-2}\,{\rm s}^{-1} \,  {\rm MeV}^{-1}
     \left(\frac{\langle P_{\nu N_1}\rangle_{\rm all-sky}}{0.35}\right)\, .
    \label{eq:fluxbenchmark}
\end{align}

For the $N_1\,\text{Xe}\to N_2\,\text{Xe}$ process, we consider a model based on a gauged $U(1)_B$, under which $N_1$ and $N_2$ are charged. The vanilla realization of such a scenario with a single heavy neutral lepton state was discussed in Refs.~\cite{Pospelov:2011ha,Pospelov:2012gm,Batell:2014yra,Kopp:2014fha,Berryman:2018jxt}. The relevant part of the Lagrangian is
\begin{align}
    \mathcal L_{\rm int}  \supset i g_{12} Z'_\mu \overline N_2\gamma^\mu N_1 + \frac{g_B}{3} Z'_\mu \sum_q \overline q\gamma^\mu q\,,
    \label{eq:vectorL}
\end{align}
and we take $N_1$ and $N_2$ to be Majorana fermions.

The vector boson, $Z'$, couples with equal strength to protons and neutrons, allowing the process $N_1\,\text{Xe}\to N_2\,\text{Xe}$ to occur, with the nucleus remaining in its ground state. The differential cross section for this process can be expressed as~\cite{Chang:2020jwl,Chao:2021bvq,Candela:2024ljb}
\begin{align}
    \frac{d\sigma_\text{Xe}}{dT}
     = &\, \frac{g_{12}^2g_B^2 A^2 m_\text{Xe}\,|F_\text{Xe}(Q)|^2}
    {4\pi(m_{Z'}^2+2m_AT)^2}
       \label{eq:vectords} \\
    & \! \times \!\left[
        2-\frac{2T}{E}+\frac{T^2-m_AT}{E^2}
        -\frac{M_2^2}{2E^2}\!
        \left( \! 1+\frac{2E-T}{m_\text{Xe}}\right)\!\right], \nonumber
\end{align}
where $F_A(Q)$ is the Helm form factor~\cite{Helm:1956zz} and $A$ denotes the total number of nucleons in the xenon nucleus.

\begin{figure}
    \centering
    \includegraphics[width=1\linewidth]{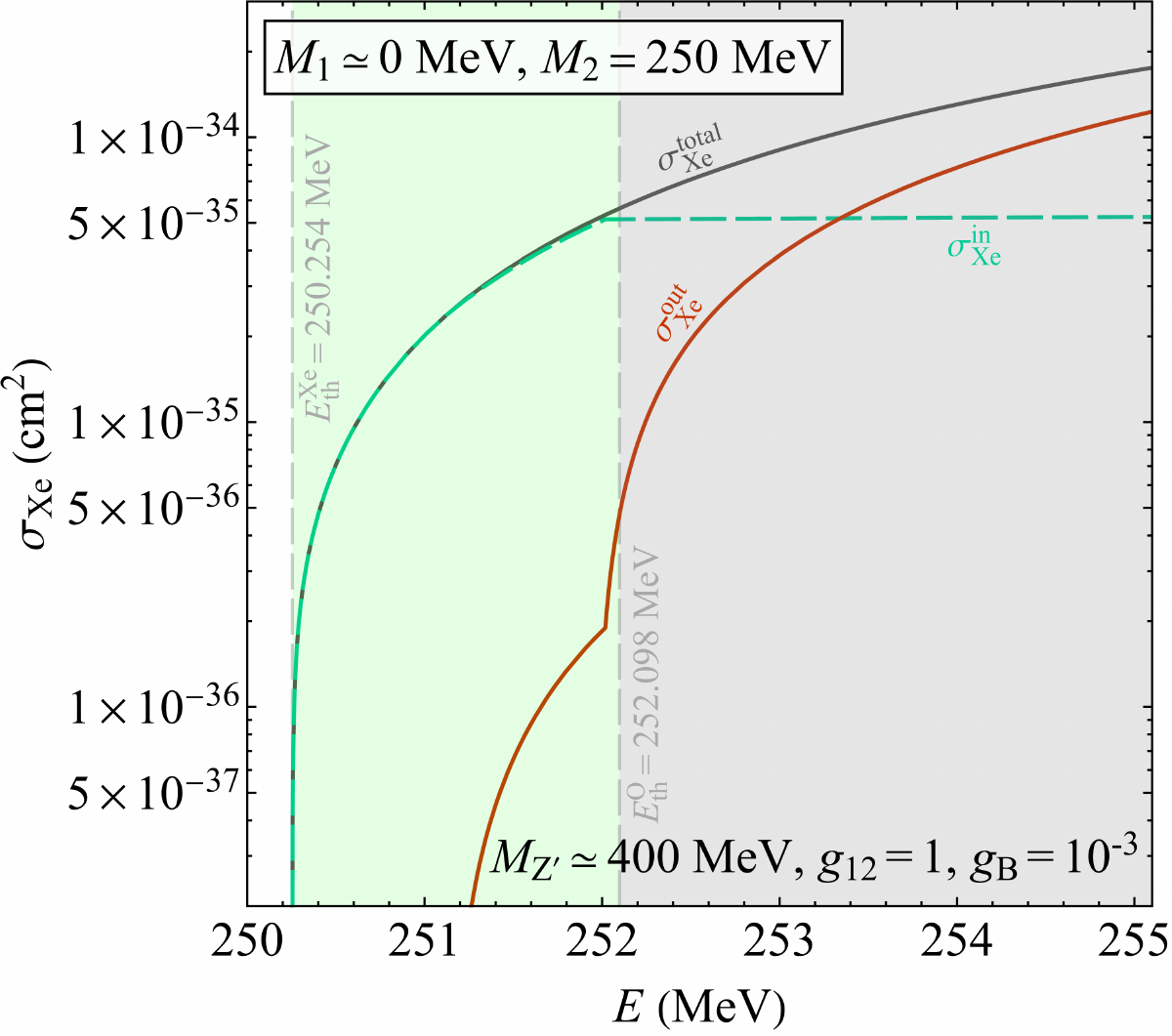}
    \caption{$N_1\,\text{Xe}\to N_2\,\text{Xe}$ scattering cross section as a function of the incident $N_1$ energy for $M_1\simeq0$ MeV, $M_2=250$~MeV, $M_{Z'}\simeq400$~MeV, $g_{12}=1$, and $g_B=10^{-3}$. The gray curve shows the total cross section, while $\sigma_{\rm Xe}^{\rm in}$ and $\sigma_{\rm Xe}^{\rm out}$ denote the contributions from recoils inside and outside the $200-300$~keV recoil window, respectively. The green region marks the interval between the xenon and oxygen scattering thresholds.
}
    \label{fig:sigmainout}
\end{figure}

Since the LZ event lies in a specific recoil energy range, we can separate the contributions inside and outside the $200-300$~keV window. We therefore define
\begin{align}
    \sigma_\text{Xe}^{\rm in}
    &=
    \int_{200~{\rm keV}}^
         {{300~{\rm keV}}}
    dT\,\frac{d\sigma_\text{Xe}}{dT} \, , \quad
    \sigma_\text{Xe}^{\rm out} = \sigma_\text{Xe}^{\rm total}-\sigma_\text{Xe}^{\rm in}\, ,
\end{align} 
where $\sigma_\text{Xe}^{\rm total}$ is obtained by integrating \cref{eq:vectords} over the entire kinematically allowed region, set by the nuclear recoil energy boundaries from \cref{eq:endpoints}.
The predicted number of events at LZ is determined by $\sigma_{\rm Xe}^{\rm in}$ rather than the total scattering cross section. On the other hand, $\sigma_{\rm Xe}^{\rm out}$ corresponds to scattering with recoil energies outside the observed signal region. In \cref{fig:sigmainout}, we present $\sigma_{\rm Xe}^\text{total}$, $\sigma_{\rm Xe}^\text{in}$, and $\sigma_{\rm Xe}^\text{out}$ for one particular benchmark point of the considered model. Close to $E_\text{th}^{\text{Xe}}$, for $M_2 =250$~MeV, the allowed recoil energies are predominantly within the $200-300$~keV window, namely $\sigma_{\rm Xe}^{\rm in}\simeq\sigma_{\rm Xe}^{\rm tot}$. As the incident energy increases, an additional recoil window opens outside this range, and $\sigma_{\rm Xe}^{\rm out}$ eventually takes over.
This, however, happens only at $N_1$ energies for which the flux must be highly suppressed (gray region), in order to avoid signals in large neutrino detectors.

The expected number of events in the recoil interval of interest is given by
\begin{align}
   N_\text{LZ}
    = \mathcal{E}_\text{Xe} \!
    \int_{E_i}^
         {E_f} 
    \!\! dE \, \frac{d\Phi_1}{dE} \int_{200~{\rm keV}}^{300~{\rm keV}} \! \! \! dT\, \epsilon_{\text{LZ}} (T)\frac{d\sigma_A}{dT}(E,T)\,,
    \label{eq:vectorsigmaavg}
\end{align}
where we take $E_i \equiv E_{\rm th}^{\rm Xe}(M_2)+0.5$~MeV and $E_f \equiv E_{\rm th}^{\rm Xe}(M_2)+1.5$~MeV, and we also take $\Phi_1$ to be constant over that energy range.
For $E_{\rm th}^{\rm Xe}(M_2)$, see \cref{eq:Eth}. For the nuclear recoil signal efficiency, $\epsilon_{\text{LZ}}$, we adopt values from Ref.~\cite{LZ:2026axp}, which are typically above 90\% for recoil energy $\lesssim 250$ keV. With the $2.84$ tonne-year LZ exposure~\cite{LZ:2026axp} and the $N_1$ flux from \cref{eq:fluxbenchmark}, we find that the process $N_1\,\mathrm{Xe}\to N_2\,\mathrm{Xe}$ requires a cross section of $\mathcal{O}(10^{-34})\,\text{cm}^2$ in order to obtain $\mathcal{O}(1)$ event in LZ in the recoil-energy range of interest. This cross section is $4$ to $5$ orders of magnitude larger than the Standard Model neutrino cross section at $E=250$~MeV. This motivates our choice of the model in \cref{eq:vectorL}, which allows the sterile neutrino to have an interaction strength much larger than that of the weak interaction \cite{Pospelov:2011ha,Dror:2017ehi,Brdar:2025azm}. In this regard, the required interaction strength also provides an additional reason to avoid the $\nu\to N$ upscattering scenario considered in \cite{Jeesun:2026vzo}: if the scattering proceeds through the neutrino interaction with quarks, the respective interaction strength cannot be several orders of magnitude larger than the Fermi constant. In our realization, by contrast, the scattering involves only heavy neutral leptons, $N_1$ and $N_2$, which interact with quarks through a $Z'$ mediator.

\begin{figure*}
    \centering
    \includegraphics[width=0.326\linewidth]{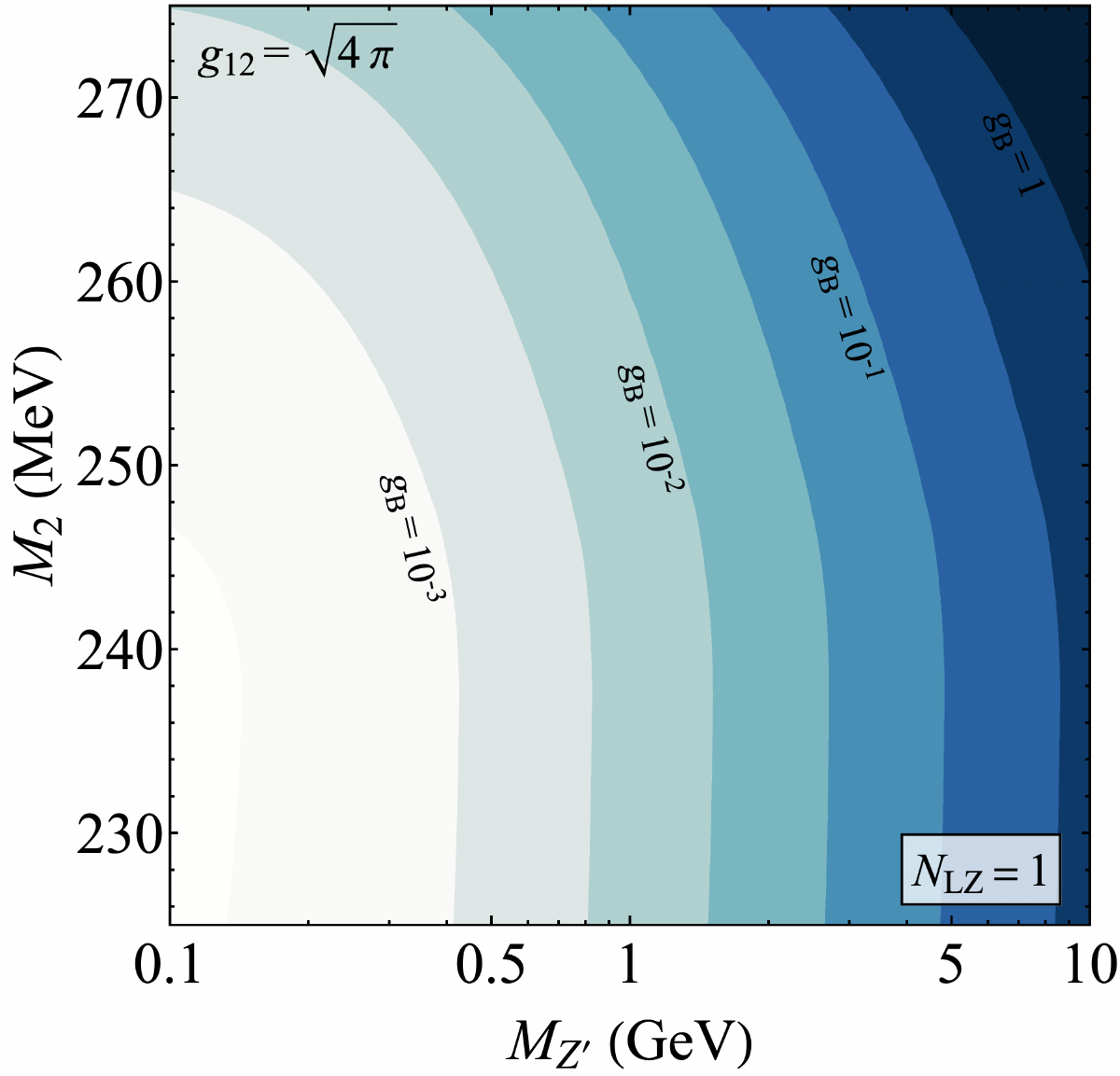}
    \includegraphics[width=0.33\linewidth]{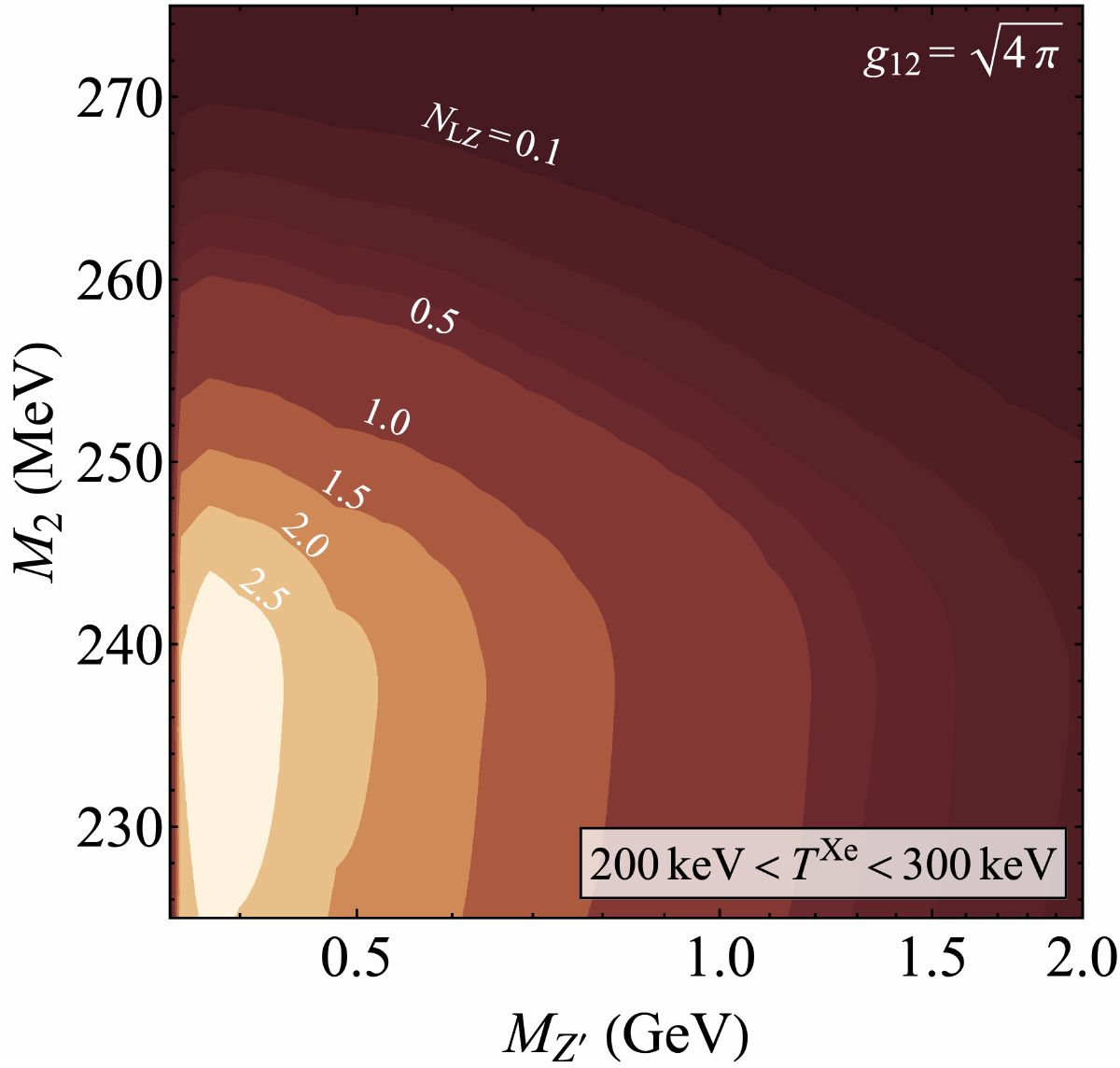}
    \includegraphics[width=0.33\linewidth]{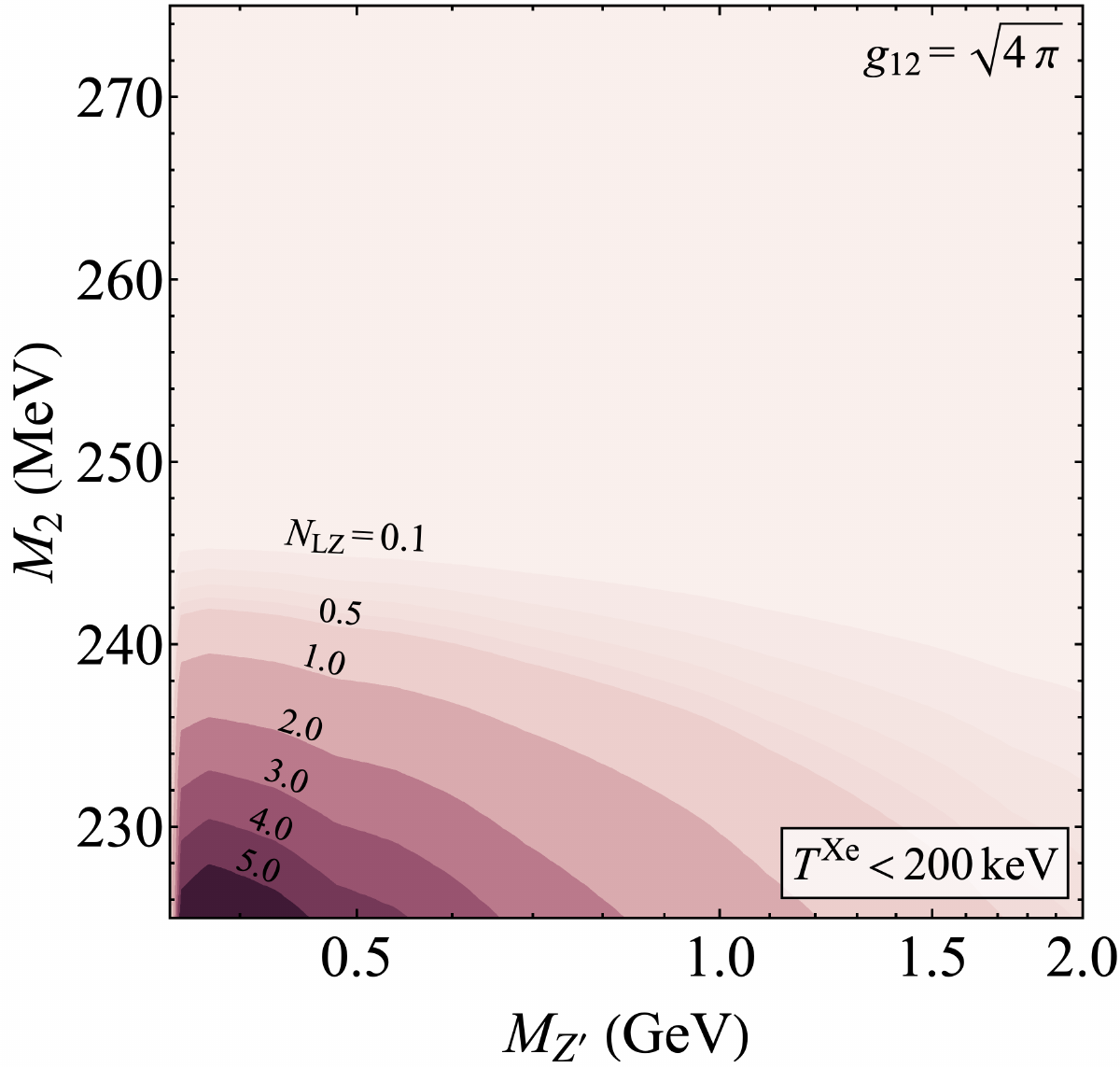}
    \caption{\textit{Left:} The value of the gauge coupling $g_B$ required to obtain one event at LZ, within the recoil window of $T^{\rm Xe} = (200 -300)$~keV, as a function of the mediator mass $M_{Z'}$ and $M_2$, for a fixed value of $g_{12}=\sqrt{4\pi}$. \textit{Middle:} Expected number of LZ events in the $(M_{Z'},M_2)$ plane for $g_{12}=\sqrt{4\pi}$, with $g_B$ fixed to the upper limit from Ref.~\cite{Dror:2017ehi}, at each $M_{Z'}$. \textit{Right:} Expected number of events outside the $(200 - 300)$~keV window, in the $(M_{Z'},M_2)$ plane for $g_{12}=\sqrt{4\pi}$, with $g_B$ treated in the same way as in the middle panel.}
    \label{fig:gbaryonic}
\end{figure*}

In \cref{fig:gbaryonic}, we present the number of predicted events at LZ as a function of $M_{Z'}$ and $M_2$. In the left panel, we show the values of $g_B$ required to obtain a single event in the $200-300$~keV recoil energy window at LZ, fixing $g_{12}=\sqrt{4\pi}$ at the perturbative limit. The required $g_B$ increases with $M_{Z'}$ to compensate for the decrease in the cross section at larger $M_{Z'}$. The decreasing LZ efficiency above recoil energies of $\sim250$~keV suppresses the event rate at larger $M_2$~\cite{LZ:2026axp}, requiring correspondingly larger values of $g_B$ to obtain an expected event count of $N_{\rm LZ} = 1$.

In the middle panel, we fix $g_B$ to its upper limit for a given $M_{Z'}$, as determined in Ref.~\cite{Dror:2017ehi},\footnote{Direct probes of the baryonic interaction generally give weaker bounds~\cite{Tulin:2014tya,Foguel:2022ppx}.} and present the number of expected LZ events, again for $g_{12}=\sqrt{4\pi}$. The $N_{\rm LZ}\geq 1$ region therefore directly identifies the region of parameter space that is consistent with existing constraints and capable of accounting for the observed LZ event. For instance, $M_{Z'}\simeq 0.5$~GeV and $M_2\simeq 250$~MeV lie on this  contour. From the left panel, we find the corresponding gauge coupling to be $g_B\simeq 10^{-3}$. Quantifying the interaction strength as $G_B\equiv g_{12}g_B/M_{Z'}^2$, we therefore find that, for this benchmark point, $G_B \simeq \mathcal{O}(10^3)\, G_F$ is required to account for the LZ event.

In the right panel of \cref{fig:gbaryonic}, we present the number of LZ events outside the $200-300$~keV recoil-energy window, again for $g_{12}=\sqrt{4\pi}$.
We do not include the higher-energy recoil contribution, since the reported LZ efficiency reaches zero above $300$~keV~\cite{LZ:2026axp}. Probing this region would require a dedicated high-energy sideband search~\cite{Rodd:2026tyn,Dent:2026bji,Langhoff:2026ujr}. 
For the aforementioned benchmark point ($M_{Z'}=0.5$~GeV, $M_2=250$~MeV), we obtain $\lesssim\mathcal{O}(0.1)$ events at low recoil, consistent with the absence of such events at LZ. This is essentially by construction since, as elaborated in \cref{sec:N1-N2}, and in particular in \cref{eq:Tstar}, a threshold recoil of $250$~keV corresponds to $M_2\simeq250$~MeV. For smaller $M_2$, the threshold recoil shifts to lower energies, and recoils below $200$~keV are easier to access.

\section{Summary and Conclusions}
\label{sec:conclusion}
\noindent
The recent observation of a $\sim 250$~keV nuclear recoil event by the LZ Collaboration may turn out to be the first signal of dark matter in a direct-detection experiment. In this work, however, we take a complementary approach and investigate whether the observed event can instead be explained by atmospheric neutrinos. The key challenge is that the broad atmospheric neutrino spectrum would also induce a large number of events in neutrino detectors, where no corresponding signal has been observed to date.

We have shown that this problem can be avoided if the atmospheric neutrino flux first produces a narrow spectrum of a new state $N_1$, followed by the process $N_1\,\text{Xe}\to N_2\,\text{Xe}$. For $M_2\simeq250$~MeV and $M_1\ll M_2$, the xenon and oxygen thresholds are separated by about $1.8$~MeV. An $N_1$ spectrum confined to this narrow energy interval can therefore scatter off xenon while remaining below the corresponding thresholds for scattering off the lighter nuclei in neutrino detectors. 

Such a narrow spectrum can arise naturally through resonant $\nu\to N_1$ conversion. We find that a conventional matter-induced resonance is less suitable for this purpose, since suppressing its high-energy tail requires increasingly smaller mixing angles, which reduce the flux of $N_1$. Parametric resonance in an ultralight scalar background provides a more effective alternative. The width of the resonance is primarily controlled by the dark matter velocity distribution, allowing substantial conversion within the desired energy interval while strongly suppressing the flux at higher energies. We find an all-sky-averaged $\nu \to N_1$ conversion probability of approximately $0.35$ over a 1 MeV window, generating an $N_1$ flux comparable to the atmospheric neutrino flux in this energy range.

The $N_1\,\text{Xe}\to N_2\,\text{Xe}$ scattering is realized within a gauged $U(1)_B$ under which both $N_1$ and $N_2$ are charged. In such a framework, interaction strengths of $\mathcal{O}(10^2$--$10^3)\,G_F$ are not excluded by existing constraints, and we find that an $\mathcal{O}(1)$ event at LZ can be realized in this regime.

In summary, the mechanism we propose in this work to explain the LZ event with neutrinos is twofold, being the combination of $(i)$ resonant $\nu\to N_1$ conversion, and $(ii)$ endothermic $N_1\to N_2$ scattering.
Together, these ingredients allow for a xenon recoil in the $\sim250$~keV range while keeping the corresponding scattering channels in large neutrino detectors kinematically inaccessible. Future data from xenon-based direct-detection experiments, as well as upcoming large-scale neutrino experiments using argon, such as DUNE, may test this scenario and shed light on the origin of the LZ event.

\section*{Acknowledgements}
We would like to thank Mom Chatterjee for illustrating the schematic diagram~(\cref{fig:schematic}).  The work of VB is supported by the United States Department of Energy Grant No. DE-SC0016013.

\bibliographystyle{JHEP}
\bibliography{lzcite}

\end{document}